\documentclass[twocolumn,10pt,a4paper,sort&compress,aps,pre,showpacs,superscriptaddress]{revtex4-2}

\usepackage[english]{babel}

\usepackage[caption=false]{subfig}
\addto\captionsenglish{}

\usepackage{bm}
\usepackage{overpic}
\usepackage{amsmath}
\usepackage{amssymb}
\usepackage{color}
\newcommand \bl{\color{black}}
\newcommand \rd{\color{black}}

\newcommand \bmOmega{\boldsymbol{{\it \Omega}}}
\newcommand \itOmega{{\it \Omega}}
\newcommand \itGamma{{\it \Gamma}}
\newcommand \itSigma{{\it \Sigma}}

\begin{document}
\title{Jeffery's Orbits and Microswimmers in Flows: A Theoretical Review}

\author{Kenta Ishimoto}
\email{ishimoto@kurims.kyoto-u.ac.jp}
\affiliation{Research Institute for Mathematical Sciences, Kyoto University, Kyoto 606-8502, Japan}

\date{\today}

\begin{abstract}
In this review, we provide a theoretical introduction to Jeffery's equations for the orientation dynamics of an axisymmetric object in a flow at low Reynolds number, and review recent theoretical extensions and applications to the motions of self-propelled particles, so-called microswimmers, in external flows. Bacteria colonize human organs and medical devices even with flowing fluid, microalgae occasionally cause huge harmful toxic blooms in lakes and oceans, and recent artificial microrobots can migrate in flows generated in well-designed microfluidic chambers. The Jeffery equations, a simple set of ordinary differential equation, provide a useful building block in modeling, analyzing, and understanding these microswimmer dynamics in a flow current, in particular when incorporating the impact of the swimmer shape since the equations contain a shape parameter as a single scalar, known as the Bretherton parameter. The particle orientation forms a closed orbit when situated in a simple shear, and this non-uniform periodic rotational motion, referred to as {\it Jeffery's} orbits, is due to a constant of motion in the non-linear equation. After providing a theoretical introduction to microswimmer hydrodynamics and a derivation of the Jeffery equations, we discuss possible extensions to more general shapes, including those with rapid deformation. In the latter part of this review, simple mathematical models of microswimmers in different types of flow fields are described, with a focus on constants of motion and their relation to periodic motions in phase space, together with a breakdown of the degenerate orbits, to discuss the stable, unstable, and chaotic dynamics of the system. The discussion in this paper will provide a comprehensive theoretical foundation for Jeffery's orbits and will be useful to understand the motions of microswimmers under various flows and to analyze and design computer and laboratory experiments, as well as in the active matter and artificial smart swimmer studies.
\end{abstract}

\maketitle

\section{Introduction}

A century ago, the British physicist George B. Jeffery published a paper 
 \cite{jeffery1922motion} describing the slow motion of an ellipsoidal particle in a viscous fluid. According to Google Scholar metrics, as of the date of this writing, Jeffery's article has been cited more than 5000 times, and half of these citations were made after 2010. This dramatic revival of this classical study is related to the recent progress in fluid dynamics on the cellular scale,  microfluidics, and their applications to biological and medical problems, as well as microrobots and active soft material sciences \cite{bechinger2016active, nguyen2019fundamentals,lauga2020fluid, alert2020physical, gompper20202020, soto2021smart}. 
 
 For a spheroidal body (an ellipsoidal body-of-rotation), the equation of the angular evolution is called {\it Jeffery's} equation, and later Bretherton \cite{bretherton1962motion} found that the equation is also applicable to an arbitrary body of revolution.

In a simple shear, the orientation of the particle obeys periodic rotation orbits, known as {\it Jeffery's orbits}. These simple dynamics are obtained as an exact solution to the Stokes equations and are thus useful when modeling a particle in more complex flow environments. These background flows are ubiquitous in nature, as in marine turbulence and flows in saturated soils, as well as in microfluidic devices in the laboratory  \cite{guasto2012fluid, rusconi2015microbes, conrad2018confined, wheeler2019not}. The rotational dynamics of a particle become more important when the particle is active, or in other words self-propelled, because a tiny difference in the orientation eventually yields a large deviation of the position after a while.

In this article, we provide a comprehensive introduction and review of the theoretical basis of self-propelled swimming particles at a small Reynolds number, focusing on the swimming dynamics in flows in connection with the Jeffery equations.  Self-propelled particles in fluids are called {\it microswimmers}, and 
examples include swimming microorganisms, swimming cells such as spermatozoa in broad taxa \cite{velho2021bank}, and microscopic synthetic and artificial self-propelled particles \cite{elgeti2015physics, yoshinaga2017simple}.  

A sphere is the simplest model of a swimming particle, and thus a spheroid is the next-order approximation of the shape of the swimmer.  Indeed, the shapes of biological microswimmers are very diverse, from the spherical algae of {\it Volvox} to elongated bacterium and from the symmetrical shape of diatoms to largely deformed shapes, such as vigorous movement of sperm \cite{dusenbery2009living, guasto2012fluid,ryabov2021shape, naselli2021life}. To elucidate the effects of shape on the motions of microswimmers in flows, Jeffery's equations have been popularly and widely used as a theoretical building block in modeling and analyzing their complex behaviors. A notable feature is the close, periodic orbit emerging as a constant of motion in non-linear differential equations. In this review, we examine the microswimmer dynamics through constants of motion, focusing on how they are derived and when the constancy is violated.   

The response of organisms to a background flow is generally called {\it rheotaxis}, and some rheotactic behaviors in biological and artificial microswimmers are understood as hydrodynamical interactions between the swimmer shape and its surrounding flow, enabling us to understand the {\bl mechanisms} and design sorting and navigating systems. From a simple shear and quadratic Poiseuille flow in pipes to more complex swirling vortical flow and mixing flow with hyperbolic stagnation points, we discuss their theoretical developments and experimental realizations.

We start with the fluid dynamics of a self-deforming particle in Sec. \ref{sec:hyd} and provide a detailed derivation for the Jeffery equations and Jeffery's orbits in Sec. \ref{sec:jef}. 
Theoretical extensions for the Jeffery equations are discussed in Sec. \ref{sec:ext}, which contains the factors that violate periodic Jeffery orbits. 
We then move on to swimming behaviors in simple to complex background flows in Sec. \ref{sec:flow}, where we review how the Jeffery equations are used in the modeling of swimming dynamics in flows. Future perspectives and topics that are not fully covered in this review are briefly discussed in Sec. \ref{sec:disc}, after which the conclusion is provided.

\section{Microswimmer Hydrodynamics}
\label{sec:hyd}
\subsection{Governing Equations}

We start with the governing equations for the hydrodynamics of microswimmers. The fluid motion obeys the incompressible Navier-Stokes equations. The Reynolds number for the flow around a swimmer is given by $Re=\mu UL/\rho$, where $\mu$ is the viscosity and $\rho$ is the fluid mass density. The characteristic length $L$ and $U$ are typically the size and swimming speed, respectively, of the microswimmer. Many swimming organisms have a Reynolds number much less than unity, for example, $Re\sim 10^{-5}$ for bacteria such as {\it Escherichia coli}, $Re\sim 10^{-3}$ for sperm cells, and $Re\sim 10^{-1}$ for a ciliate {\it Paramecium} \cite{lauga2009hydrodynamics}, where the Stokes approximation well represents the flow dynamics. Hence, neglecting the inertia terms in the Navier-Stokes equation reduces our fluid equations to the steady Stokes equation given by 
\begin{equation}
\nabla p=\mu\nabla^2 \bm{u}
    \label{eq:M01},
\end{equation}
where $p$ is the pressure and $\bm{u}$ is the fluid velocity that satisfies the incompressible condition $\nabla\cdot\bm{u}=0$. 
The Stokes equations may be rewritten as $\partial_j\sigma_{ij}=0$ by using the Newtonian stress tensor, $\sigma_{ij}=-p\delta_{ij}+2\mu E_{ij}$.
Here, we use the Einstein summation convention for the repeated spatial indices $i,j=\{1, 2, 3\}$, $\partial_i=\partial/\partial x_i$, $\delta_{ij}$ is the Kronecker delta,  and $E_{ij}=(\partial_{i}u_{j}+\partial_{j}u_{i})/2$ is the rate-of-strain tensor.

\begin{figure}[tbp]
\begin{center}
\includegraphics[width=9cm]{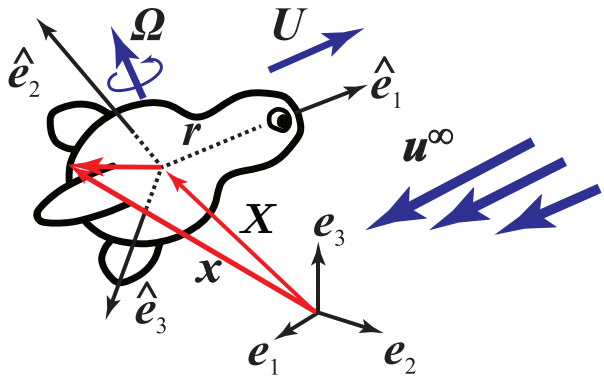}
\caption{ (Color online) Schematic of a microswimmer in a fluid. Two reference systems are considered to describe the swimming dynamics: The laboratory-fixed coordinates $\{\bm{e}_i\}$ and the body-fixed coordinates $\{\hat{\bm{e}}_i\}$, whose origin is denoted by $X_i$. The orientation of the swimmer is represented by a rotational matrix that corresponds to the two coordinate systems. Figure modified from Ishimoto \cite{ishimoto2020jeffery} under the creative commons license, http://creativecommons.org/licenses/by/4.0. {\rd (copyright 2020 by The Author(s)) }
}
\label{fig:coord}
\end{center}
\end{figure}

We now consider the equation of motion of a swimmer immersed in fluid. To specify the motion of a deforming object, we introduce two reference frames, the laboratory-fixed frame, $\{\bm{e}_i\}$, and the swimmer-fixed frame, $\{\hat{\bm{e}}_i\}$ (Fig. \ref{fig:coord}).  The linear and angular momentum conservation should hold. Again, at low Reynolds numbers, all the inertia terms are negligible and thus these momentum equations are reduced to the force and torque balance relation. The total force $\bm{F}$ and the torque $\bm{M}$ acting on the object should vanish as
\begin{equation}
F_i=\int_S \sigma_{ij}n_j\,dS{\bl =0},~M_i=\int_S \epsilon_{ijk}(x_j-X_j)\sigma_{k\ell}n_\ell\,dS=0
    \label{eq:M02},
\end{equation}
{\bl if there is no other external forces or torques. Here,} $S$ indicates the surface of the swimmer, the unit normal $\bm{n}$ is taken outwards to the surface, and $\epsilon_{ijk}$ is the Levi-Civita permutation symbol. The vector $\bm{X}$ indicates the origin of the body-fixed coordinates. At the limit of negligible inertia, replacements of the body-fixed coordinates and their origin do not affect the equations of motion \eqref{eq:M02}. This corresponds to the gauge freedom of the system, and the swimmer dynamics are determined essentially by the shape gait \cite{shapere1989geometry}.  
 To solve the swimmer dynamics, we may practically fix the body-fixed coordinates, or fix the gauge arbitrarily, although a theoretically reasonable choice is to set the origin as the center of mass \cite{ishimoto2012coordinate}. 

 We also consider the background flow  $\bm{u}^\infty$ and assume this given function also satisfies the Stokes equations. The total fluid velocity is a sum of the background flow and the disturbed flow due to the presence of the swimmer $\bm{u}^d$, that is, $\bm{u}=\bm{u}^\infty+\bm{u}^d$. In many situations in nature and the laboratory, the length scale for modulation of the background flow is larger than the size of the swimmer. Even in the turbulent ocean, the smallest length scale of the vortices is estimated at around 10--30 mm \cite{guasto2012fluid}, which is sufficiently large to approximate the local background flow to be linear in position. The background $\bm{u}^\infty$ may be regarded as the flow induced by other swimmers or external boundaries, and these hydrodynamic interactions are discussed in later sections.
 
 Now, let us consider the dynamics in the absence of a flow ($\bm{u}^\infty=\bm{0})$ and  write down the velocity of the swimmer surface $\bm{v}$ to describe the no-slip boundary conditions of the fluid equation \eqref{eq:M01}:
 \begin{equation}
     \bm{u}^d=\bm{v} ~\textrm{on}~S
     \label{eq:M03}.
 \end{equation}

 Let $\bm{a}$ be the Lagrangian label for the material point on the surface $S$. We can specify the point on the surface by using the body-fixed coordinate as $\bm{x}(\bm{a}, t)=\bm{X}+x^s_i(\bm{a}, t)\hat{\bm{e}}_i$. Its time derivative gives 
\begin{equation}
\bm{v}=\frac{d\bm{x}}{dt}=\frac{d\bm{X}}{dt}+x^s_i\frac{d\hat{\bm{e}_i}}{dt}+\frac{\partial x^s_i}{\partial t}\hat{\bm{e}_i}
    \label{eq:M04a}
\end{equation}
when the time derivative in the final term is taken with a fixed Lagrangian label.
We then define the translational velocity $\bm{U}$ and rotational velocity $\bmOmega$ as the difference between the two frames as $\bm{U}=d\bm{X}/dt$ and 
$d\hat{\bm{e}}_i/dt=\bmOmega\times\hat{\bm{e}}_i$. Introducing the deformation velocity as $\bm{u}^s=(\partial x^s_i/\partial t)\hat{\bm{e}}_i$, we may obtain the surface velocity as
\begin{equation}
\bm{v}=\bm{U}+\bmOmega\times (\bm{x}-\bm{X})+\bm{u}^s
    \label{eq:M04b}.
\end{equation}

In this article, we particularly focus on the kinematic problem, in which one solves the swimming velocity, $\bm{U}$ and $\bmOmega$, and thus the swimming trajectory under a given shape gait $x^s_i(\bm{a}, t)$. If we need to solve the shape gait as an unknown function by solving a coupling problem with the material elasticity, this is called elastohydrodynamics \cite{ishimoto2018elastohydrodynamical, kawamura2018phase, lagrone2019elastohydrodynamics, ishimoto2022self}, and the constitutive relation of the swimmer, $\sigma^{\textrm{int}}_{ij}(t)$, is required to complete the model equations. At the zero inertia limit, the Cauchy momentum equation inside the swimmer is $\partial_j\sigma^{\textrm{int}}_{ij}=0$ and by the Gauss theorem we have
\begin{equation}
    F_i=-\int_S \sigma^{\textrm{int}}_{ij}n_j\,dS=0
    \label{eq:M05},
\end{equation}
which indicates that no force is generated from the interior of the swimmer and also for the torque, regardless of the material property. Hence, the system is closed by the Stokes equations and the force- and torque-free conditions.

In summary, the governing equations in the kinematic problem of microswimmer hydrodynamics are therefore Eqs. \eqref{eq:M01} and \eqref{eq:M02} with the no-slip boundary condition $\bm{u}^d=\bm{v}$ on the surface $S$ given by Eq. \eqref{eq:M04b}. Depending on the confinement that encloses the fluid, we also need further boundary conditions on the surface of the container. If the swimmer is located in a free space, we impose $\bm{u}^d\rightarrow\bm{0}$ as $|\bm{x}-\bm{X}|\rightarrow \infty$.

\subsection{Swimming Formula}

We now consider the hydrodynamic forces and torques acting on a microswimmer and derive the swimming formula that provides us with the swimming velocities for a given shape gait. 

To do so, we first recall the {\it Lorentz reciprocal theorem}, which is a relation between the solutions to the Stokes equations with the same boundary geometry $S$ but with different boundary values \cite{masoud2019reciprocal}. Let $(u_i, \sigma_{ij})$ and $(\tilde{u}_i, \tilde{\sigma}_{ij})$ be the two distinct Stokes solutions, and from the Lorentz reciprocal theorem, these satisfy 
\begin{equation}
\int_S u_i\tilde{\sigma}_{ij} n_j dS= \int_S \tilde{u}_i \sigma_{ij} n_j dS
    \label{eq:M11},
\end{equation}
where $\bm{n}$ indicates the unit normal on the surface. 

Now we assume that the former solution $(u_i, \sigma_{ij})$ corresponds to solutions around a deforming swimmer whose surface velocity is given by Eq. \eqref{eq:M04b} in a background flow, and that the latter with the hat symbol $(\tilde{u}_i, \tilde{\sigma}_{ij})$ is chosen for a flow around a rigid body with arbitrary translational and rotational velocities, $\tilde{\bm{U}}$ and $\tilde{\bmOmega}$, respectively, in the absence of a background flow.

With the linearity of the flow equation, we may decompose the surface force on the rigid body (from the surrounding fluid) $\tilde{f}_i=-\tilde{\sigma}_{ij}n_j$, with $\bm{n}$ being the outward unit normal to the fluid region,  in the form 
\begin{equation}
    \tilde{f}_i=-\itSigma^U_{ij}\tilde{U}_j-\itSigma^\Omega_{ij}\tilde{\itOmega}_j
    \label{eq:M12a}. 
\end{equation}
The second-rank tensors, $\itSigma^U_{ij}$ and $\itSigma^\itOmega_{ij}$, are called the translational and rotational surface force resistance tensors, respectively \cite{pozrikidis1992boundary}. As seen below, these tensors are fundamental quantities to solve the swimmer dynamics from a theoretical point of view, even though standard direct numerical computations of the Stokes flow bypass these by using the boundary integral formulation \cite{pozrikidis1992boundary}. Note, however, that these tensors are functions depending only on the instantaneous shape of the swimmer when represented in the body-fixed coordinates once the fluid viscosity is fixed. 

After integrating these variables over the surface $S$, we obtain the familiar resistance tensors by introducing four $3\times3$ tensors as $K^{FU}_{ij}=\int_S \itSigma^U_{ij}dS$, $K^{F\itOmega}_{ij}=\int_S \itSigma^\itOmega_{ij}dS$, $K^{MU}_{ij}=\int_S \epsilon_{ik\ell}r_k\itSigma^U_{\ell j} dS$, and $K^{M\itOmega}_{ij}=\int_S \epsilon_{ik\ell}r_k\itSigma^\itOmega_{\ell j} dS$, where we have introduced $r_k=x_k-X_k$. These are integrated into a single  $6\times 6$ tensor, 
\begin{equation}
 \mathcal{K}=\begin{pmatrix}
     {\bf K}^{FU} & {\bf K}^{F\itOmega},\\
     {\bf K}^{MU} & {\bf K}^{M\itOmega}
 \end{pmatrix}
 \label{eq:M12},
\end{equation}
 called the {\it grand resistance tensor}, which is symmetric and positive-definite \cite{happel2012low}. Hence, one can define its inverse, $\mathcal{M}=\mathcal{K}^{-1}$, called the grand mobility tensor. We then introduce 6-dimensional velocity and force vectors as
\begin{equation}
 \mathcal{U}=\begin{pmatrix}
     \bm{U} \\ \bmOmega
 \end{pmatrix},~
  \hat{\mathcal{U}}=\begin{pmatrix}
     \tilde{\bm{U}} \\ \tilde{\bmOmega}
 \end{pmatrix},~
  \mathcal{F}=\begin{pmatrix}
     \bm{F} \\ \bm{M}
 \end{pmatrix},~
   \hat{\mathcal{F}}=\begin{pmatrix}
     \tilde{\bm{F}} \\ \tilde{\bm{M}}
 \end{pmatrix}
 \label{eq:M12}.
\end{equation}

Plugging this representation into Eq. \eqref{eq:M11}, we may rewrite it in the form \cite{yariv2006self, ishimoto2012coordinate, ishimoto2020helicoidal} 
\begin{equation}
\mathcal{U}_a\hat{\mathcal{F}}_a+\hat{\mathcal{U}}_a\mathcal{F}^{\textrm{prop}}_a+\hat{\mathcal{U}}_a\mathcal{F}^{\infty}_a=\hat{\mathcal{U}}_a\mathcal{F}_a
    \label{eq:M14},
\end{equation}
{\bl where we use} the indices $a, b, \cdots,\in \{1, 2, \cdots, 6 \}$ to distinguish from the indices $i, j, k, \cdots, \in \{1, 2, 3\}$ for the spatial labels.
Here, we have written
\begin{equation}
  \mathcal{F}^\textrm{prop}=\begin{pmatrix}
     \bm{F}^{\textrm{prop}} \\ \bm{M}^{\textrm{prop}}
 \end{pmatrix},~ \mathcal{F}^{\infty}=\begin{pmatrix}
     \bm{F}^{\infty} \\ \bm{M}^{\infty}
 \end{pmatrix}
 \label{eq:M15a},
\end{equation}
and these are explicitly given by
\begin{equation}
F^{\textrm{prop}}_j=-\int_S u'_i \itSigma^{U}_{ij}dS,~
M^{\textrm{prop}}_j=-\int_S u'_i \itSigma^{\itOmega}_{ij}dS
    \label{eq:M15b},
\end{equation}
and 
\begin{equation}
F^{\infty}_j=\int_S u^\infty_i \itSigma^{U}_{ij}dS,~
M^{\infty}_j=\int_S u^\infty_i \itSigma^{\itOmega}_{ij}dS
    \label{eq:M15c}.
\end{equation}
Eq. \eqref{eq:M14} is an integral formula first derived by Stone \& Samuel \cite{stone1996propulsion}.
Here, the force and torque for a rigid body are simply written as 
$\hat{\mathcal{F}}_a=-\mathcal{K}_{ab}\hat{\mathcal{U}}_b$, and 
 Eq. \eqref{eq:M14} may be further reduced
to 
\begin{equation}
    \mathcal{F}_a=-\mathcal{K}_{ab}\mathcal{U}_b+\mathcal{F}^{\textrm{prop}}_a+\mathcal{F}^{\infty}_a
    \label{eq:M16}
\end{equation}
by noting that $\hat{\mathcal{U}}_a$ is taken arbitrarily. For a swimmer under the force- and torque-free conditions \eqref{eq:M02}, that is, $\mathcal{F}_i=0$, the swimming formula is then simply given by
\begin{equation}
\mathcal{U}_a=\mathcal{M}_{ab}\,(\mathcal{F}^{\textrm{prop}}_b+\mathcal{F}^{\infty}_b)
    \label{eq:M17},
\end{equation}
from which one can calculate the swimming velocities for a given shape gait and background flow.
 With the swimming formula \eqref{eq:M17}, one may be able to explicitly demonstrate  {\it Purcell's scallop theorem}\cite{purcell1977life}, which states that a microswimmer cannot generate net locomotion by reciprocal motion. The reciprocal motion should be defined as the time-reversal deformation using the body-fixed surface position.  Note, however, that 
 the surface velocity $\bm{u}^s(\bm{a},t)=(\partial x^s_i/\partial t)\hat{\bm{e}}_i$ depends on the instantaneous orientation of the swimmer, and we thus need representations of the resistance tensors in the body-fixed coordinates to provide a formal proof for the scallop theorem \cite{ishimoto2012coordinate}.

\subsection{Squirmer}

\begin{figure}[htbp]
\begin{center}
\begin{overpic}[width=8.5cm]{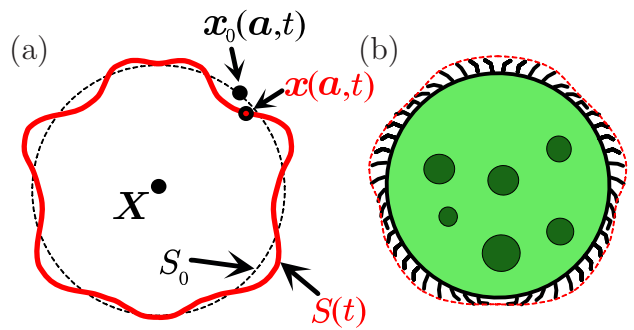}
\put(1,45){\large (a)}
\put(55,45){\large (b)}
\end{overpic}
\caption{ (Color online) Schematic of a slightly deforming microswimmer. (a) Sphere with slight deformation. The instantaneous shape shown with a thick red curve is denoted by $S(t)$, whereas the original unperturbed sphere is shown as a dotted black circle denoted by $S_0$. The point on the surface with the Lagrangian label $\bm{a}$ moves from $\bm{x}_0(\bm{a}, t)$ to $\bm{x}(\bm{a}, t)$. (b) Schematic of a swimming ciliate whose shape is well described by the envelope shown with a broken red curve.}
\label{fig:squirm}
\end{center}
\end{figure}

In this subsection, we consider a specific swimmer model, known as the {\it squirmer} {\rd (Fig. \ref{fig:squirm}}. The squirmer was first proposed by Lighthill \cite{lighthill1952squirming} as a slightly deforming sphere and later {\bl corrected and} applied by Blake to analyze a swimming ciliate \cite{blake1971spherical}. Presently, however, the squirmer refers to a mathematical model of a rigid body with a surface slip velocity instead of actual deformation \cite{pedley2016squirmers}.
Due to the model's simplicity, the squirmer has been extensively studied in the past few decades to elucidate universal features of microswimmers and their hydrodynamic interactions, as well as to model the behaviors of specific microorganisms, such as ciliates and volvox and artificial swimming colloids \cite{magar2003nutrient, ishikawa2006hydrodynamic, michelin2010efficiency, ishimoto2013squirmer, lauga2014locomotion, uspal2015rheotaxis, goldstein2015green, pedley2016squirmers, omori2020swimming, qi2022emergence}.

Let us consider a general deforming object, express a position vector on its surface as $\bm{x}(\bm{a}, t)=\bm{X}+x^s_i(\bm{a}, t)\hat{\bm{e}}_i$, and write its deformation as $x^s_i(\bm{a},t)=\epsilon \alpha_i(\bm{a}, t)$, where $\epsilon$ denotes the magnitude of the deformation and $\alpha_i$ is taken as $|\alpha_i|=O(1)$.
We analyze the swimming dynamics for a small deformation, specifically $O(\epsilon)\ll 1$.

Our primary aim is to calculate the surface force resistance tensors \eqref{eq:M12a} to solve the hydrodynamic force/torque or to apply the swimming formula \eqref{eq:M17}. Because it is impractical to obtain these quantities in an analytical form for a general deforming object, we instead construct an equivalent but simpler boundary condition in which the Stokes solutions remain the same. To do so, we introduce the reference surface $S_0(t)$, which denotes the surface in the absence of deformation, and this is obtained by setting $\epsilon=0$ in the expression for $x^s_i$. By Eq. \eqref{eq:M04a}, the surface deformation velocity is given by $\bm{u}'=\epsilon \dot{\alpha}_i\hat{\bm{e}}_i$, where the dot indicates the Lagrangian time derivative. 

Instead of the no-slip boundary conditions on $S(t)$, we consider a slip surface velocity $\bm{u}'_0=A_i(\bm{a},t)\hat{\bm{e}}_i$ on $S_0(t)$, and determine the coefficient $A_i$ to ensure that the fluid velocity behaves properly on $S(t)$. We expand the swimming velocities and the coefficient $A_i(\bm{a},t)$ as
$\mathcal{U}_i=\epsilon\mathcal{U}^{(1)}_i+\epsilon^2\mathcal{U}^{(2)}_i+\cdots$ and $A_i=\epsilon A^{(1)}_i+\epsilon^2 A^{(2)}_i+\cdots$. The velocity fields are then also expanded around the point $\bm{x
}_0$ so that
\begin{equation}
\bm{u}(\bm{x},t)=\bm{u}(\bm{x}_0, t)+\bm{s}\cdot\nabla\bm{u}+O(\epsilon^2)
    \label{eq:M21},
\end{equation}
where $\bm{s}=\bm{x}-\bm{x}_0=O(\epsilon)$. We substitute the two expressions of the boundary conditions on $S(t)$ and $S_0(t)$ into Eq. \eqref{eq:M21}, and equating the $O(\epsilon)$ contributions in both cases yields the simple relation $A^{(1)}_i=\dot{\alpha}_i$. By setting $\bm{u}'_0=\epsilon A^{(1)}_i\hat{\bm{e}}_i+O(\epsilon^2)$, the leading-order contribution to the propulsive force is therefore obtained as
\begin{equation}
F^{\textrm{prop}}_j=-\int_{S_0} u'_{0i}
\itSigma^{U}_{0ij} dS+O(\epsilon^2)
\label{eq:M22},
\end{equation}
where $\Sigma^U_{0ij}$ indicates the surface force tensor evaluated at the reference shape $S_0$. The propulsive torque is also written only in terms of the surface force tensor at $S_0$. From the swimming formula, it is therefore found that the leading-order velocities $\bm{U}^{(1)}$ and $\bmOmega^{(1)}$ are the same as those of a swimmer with a slip velocity on its reference surface $S_0$. In other words, for a small deformation, a deforming swimmer is well approximated by a rigid body with surface slip velocity, that is, a simple squirmer.

The squirmer model has been extensively studied, in particular to study the hydrodynamic interactions between the external boundaries and between the cells. In particular, the symmetric, traceless part of the first force moment $S_{ij}$, called the {\it stresslet}, determines the leading-order contribution to the far-field flow around a swimmer under a force- and torque-free condition. The stresslet is defined as
\begin{equation}
S_{ij}=\frac{1}{2}\int_S [f_ix_j+f_jx_i+\frac{2}{3}f_kx_k\delta_{ij}+2\mu(u_i{\bl n_j}+u_jn_i)]\,dS
    \label{eq:M23},
\end{equation}
where $f_i=-\sigma_{ij}n_j$ is the surface force on the swimmer \cite{batchelor1970stress, ishikawa2006hydrodynamic}. For an axisymmetrical swimmer, the stresslet is represented by a single parameter $P$ as $S_{ij}=P(d_id_j-\frac{1}{3}\delta_{ij})$ with $\bm{d}$ denoting the director vector along the axis of symmetry.
The swimmer pattern is often categorized via the sign of $P$ and a swimmer with $P>0$, $P<0$, or $P=0$ is called a pusher, puller, or neutral swimmer, respectively. A pusher swimmer is typically found in a cell with an appendage at its rear, such as bacteria and sperm cells \cite{drescher2011fluid, ishimoto2017coarse}. A puller often has flagella at the front of the cell and a {\it Chlamydomonas} cell is a popular example \cite{drescher2010direct, guasto2010oscillatory}. A cell with uniform distribution of propulsive appendages, such as a ciliate, is well approximated as a neutral swimmer \cite{ohmura2018simple}.

\section{Jeffery's Equation}
\label{sec:jef}

In this section, we give a derivation of the Jeffery equation, considering a general rigid body or a squirmer swimmer in a linear background flow. 

\subsection{Linear Background Flow}

As explained in the previous section, the background flow in nature and the laboratory is often locally approximated by a linear flow, which satisfies the Stokes equations. Let us expand the background flow around the swimmer position $\bm{X}$. By rewriting the local background linear and angular velocities as $\bm{U}^\infty=\bm{u}^\infty_i(\bm{X})$ and $\bmOmega^\infty=(1/2)\nabla\times \bm{u}^\infty_i(\bm{X})$, we obtain up to the first order of the relative position $r_i=x_i-X_i$ as
\begin{equation}
u^\infty_i(\bm{x})=U^\infty_i+\epsilon_{ijk}\itOmega^\infty_j r_k+E^\infty_{ij}r_j+O(|\bm{r}|^2)
    \label{eq:J01},
\end{equation}
where $E^\infty_{ij}=(\partial_j u^\infty_i(\bm{X})+\partial_i u^\infty_j(\bm{X}))/2$ is the local rate-of-strain tensor at the position of the swimmer.

We then substitute the linear expression \eqref{eq:J01} into Eq. \eqref{eq:M15c} to obtain the flow-induced force and torque as
\begin{eqnarray}
F^\infty_i&=&K^{FU}_{ij}U^\infty_j+K^{F\itOmega}_{ij}\itOmega^\infty_j+\itGamma^F_{ijk}E^\infty_{jk}
    \label{eq:J02a} \\
M^\infty_i&=&K^{MU}_{ij}U^\infty_j+K^{M\itOmega}_{ij}\itOmega^\infty_j+\itGamma^M_{ijk}E^\infty_{jk}
    \label{eq:J02b}, 
\end{eqnarray}
where the third-rank tensors $\itGamma^F_{ijk}$ and $\itGamma^M_{ijk}$ are respectively called the force-shear tensor and torque-shear tensor, defined as
\begin{equation}
\itGamma^F_{ijk}=\int_S \itSigma^U_{ji}r_k\,dS~,~~
\itGamma^M_{ijk}=\int_S \itSigma^\itOmega_{ji}r_k\,dS
    \label{eq:J02c}.
\end{equation}
With the symmetric property of the rate-of-strain tensor, these third-rank tensors are symmetric under the exchanges of the last two subscripts,  namely $\itGamma^F_{ijk}=\itGamma^F_{ikj}$.

These flow-induced force and torque equations \eqref{eq:J02a} and \eqref{eq:J02b} are further simplified as
\begin{equation}
\mathcal{F}^{\infty}_a=\mathcal{K}_{ab}\mathcal{U}^\infty_b+\mathcal{G}_{ajk}E^\infty_{jk}
\label{eq:J03}.
\end{equation}
Here, we have introduced the 6-dimensional background velocity vector $\mathcal{U}^\infty$, $6\times 3\times 3$ tensor $\mathcal{G}_{a jk}$, given by
\begin{equation}
\mathcal{U}^\infty_{i}=
    U^\infty_i,  ~\mathcal{U}^\infty_{i+3}=\itOmega^\infty_i,~
\mathcal{G}_{ijk}=\itGamma^U_{ijk},~
\mathcal{G}_{i+3,jk}=\itGamma^\itOmega_{ijk}
    \label{eq:J03b}.
\end{equation}
The $6\times 3\times 3$ tensor $\mathcal{G}_{a jk}$ is only a function of the instantaneous shape when represented in the body-fixed coordinates if the viscosity $\mu$ is a fixed constant.

\subsection{Jeffery's Equation}

To derive the Jeffery equations, let us first consider the symmetry of the shape of a swimmer, which in turn determines the swimming dynamics. The number of constants appearing as entries of the tensors is 51 in total, but the symmetry of the object can dramatically reduce this number. 

Now we consider the angular dynamics of a body of revolution under a linear shear flow to derive the Jeffery equation. We denote the axis of rotation by a unit vector $\bm{d}$ and set it as $\bm{d}=\hat{\bm{e}}_1$. Because the body of revolution has an axisymmetric nature, we may represent the resistance tensors represented in the laboratory frame as
\begin{eqnarray}
K^{FU}_{ij}&=&K^U_{||}d_id_j+K^U_{\perp}(\delta_{ij}-d_id_j), \label{eq:J10a} \\
K^{M\itOmega}_{ij}&=&K^\itOmega_{||}d_id_j+K^\itOmega_{\perp}(\delta_{ij}-d_id_j) \label{eq:J10b} ,
\end{eqnarray}
where the coefficients $K^U_{||}, K^U_{\perp}, K^\itOmega_{||}$, and $K^\itOmega_{\perp}$ are positive constants depending on the shape of the object.
The coupling terms can be taken as zero, that is, $
K^{F\itOmega}_{ij}=K^{MF}_{ij}=0$, if the origin of the body-fixed frame is taken to coincide with the hydrodynamic center of resistance \cite{happel2012low}.

Similar symmetric arguments enable us to reduce the representation of the shear-torque resistance tensor in the laboratory frame as 
\begin{equation}
\itGamma^M_{ijk}=\itGamma({\bl \epsilon_{ik\ell}d_j+\epsilon_{jk\ell}d_i})d_\ell
\label{eq:J12a}.
\end{equation}

From the swimming formula \eqref{eq:M17}, we  use the diagonal representation of the grand resistance tensors \eqref{eq:J10a},  \eqref{eq:J10b}, and \eqref{eq:J12a} to obtain the angular velocity of the swimmer in the form
\begin{equation}
\itOmega_i=\itOmega^\infty_{i}-(K^{M\itOmega})^{-1}_{ij}(\itGamma^M_{{\bl jk\ell}} E^\infty_{k\ell}+M^{\textrm{prop}}_j)
    \label{eq:J12b}.
\end{equation}
With $\dot{\bm{d}}=\bmOmega\times \bm{d}$, we arrive at the Jeffery equation given by
\begin{equation}
\dot{d}_i=\epsilon_{ijk}\itOmega^\infty_jd_k+B(\delta_{ij}-d_id_j)E^\infty_{jk}d_k+\dot{d}^{\textrm{\,prop}}_i
    \label{eq:J12c},
\end{equation}
where we have introduced $B=\itGamma/K^\itOmega_\perp$ and $\dot{\bm{d}}^{\,\textrm{prop}}=\bm{M}^\textrm{prop}\times \bm{d}/K^\itOmega_\perp$. The scalar $B$, known as the Bretherton parameter, is only a factor that reflects the shape of the object. Indeed, for a spheroidal case, the resistance tensors are exactly obtained \cite{kim2013microhydrodynamics} and the Bretherton parameter is given by $B=(c^2-1)/(c^2+1)$, with $c$ being the aspect ratio of the spheroid. The Bretherton parameter becomes $B=0$ for a sphere and approaches $B\rightarrow 1$ at the slender limit and $B\rightarrow -1$ at the disk limit. Nonetheless, extreme cases are known in which the Bretherton parameter becomes $|B|>1$ \cite{bretherton1962motion, singh2013rigid}.

\begin{figure}[tbp]
\begin{center}
\begin{overpic}[width=8.5cm]{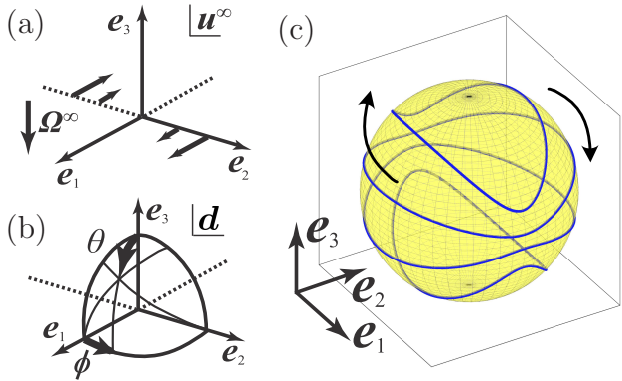}
\put(3,56){{\large (a)}}
\put(3,24){{\large (b)}}
\put(45,55){{\large (c)}}
\end{overpic}
\caption{ (Color online) Schematic of (a) a simple shear, (b) definitions of angles, and (c) sample trajectories of Jeffery's orbits. The time evolution of the orientation is illustrated by blue trajectories on a unit sphere. The plots are made for an elongated object with $B=0.9$.}
\label{fig:jeffery}
\end{center}
\end{figure}

To see the precise rotational dynamics, we then focus on the case of simple shear. Let us fix the background flow as $\bm{u}^\infty=\gamma y\bm{e}_1$, so that  the background flow possesses uniform vorticity field $\bm{\omega}^\infty=\nabla\times\bm{u}^\infty=-\bm{e}_3$ (Fig. \ref{fig:jeffery}a). 

We then rewrite Eq. \eqref{eq:J12c} by assuming that $\dot{\bm{d}}^{\,\textrm{prop}}=\bm{0}$ for simplicity. With the polar coordinates (see the schematic in  Fig. \ref{fig:jeffery}b), we have
\begin{eqnarray}
\gamma^{-1}\dot{d}_1&=&(B+1)d_2/2-Bd^2_1d_2     \label{eq:J15a} \\
\gamma^{-1}\dot{d}_2&=&(B-1)d_1/2-Bd_1d^2_2 \label{eq:J15b} \\
\gamma^{-1}\dot{d}_3&=& -Bd_1d_2d_3            \label{eq:J15c}.
\end{eqnarray}
These are further simplified by calculating $\dot{d}_3=-\dot{\theta}\sin\theta$ and $d_1\dot{d}_2-\dot{d}_1 d_2=\dot{\phi}\sin^2\theta$ as
\begin{equation}
\frac{d\theta}{dt}=\frac{\gamma}{4}B\sin2\theta\sin2\phi  ~,~~
\frac{d\phi}{dt}=-\frac{\gamma}{2}\left( 1-B\cos 2\phi\right) 
\label{eq:J16}.
\end{equation}

For a sphere ($B=0$), the object just rotates by the background angular velocity $|\bmOmega^\infty|=\gamma/2$.
The second equation is readily integrated and we obtain the periodic angular dynamics if $|B|<1$. The polar angle is determined by the azimuthal angle dynamics, and by setting the initial azimuthal angle as $\phi_0=0$, we have
\begin{eqnarray}
    \tan\theta=\frac{K c}{\sqrt{\cos^2\phi+c^2\sin^2\phi}} 
   ,~\tan\phi=-c\tan\left( \frac{\gamma t}{c+c^{-1}}\right) 
   \label{eq:J17},
\end{eqnarray}
where $K$ is a constant of the integral and acts as a constant of motion. Indeed, $K$ is determined once the initial angles are set, and $K=0$ corresponds to a rotation along the equator, while $K\rightarrow \infty$ and $K\rightarrow -\infty$ indicate spinning around the poles. {\bl Here,} $c=\sqrt{(1+B)/(1-B)}$ is therefore interpreted as an effective aspect ratio from the Bretherton parameter. The orientation vector $\bm{d}$ traces a closed orbit {\bl on the unit sphere and} determined by the initial {\bl value of $K$.} {\bl These closed orbits are formed by 
infinitely many marginally stable periodic orbits and are called degenerated.} 
The period of one rotation, 
\begin{equation}
T=\frac{2\pi}{\gamma}\left(c+\frac{1}{c} \right)=\frac{4\pi}{\gamma\sqrt{1-B^2}}
    \label{eq:J18},
\end{equation}
does not depend on $\theta$ but is determined by the shear strength and the Bretherton parameter.
This non-linear periodic motion is the Jeffery's orbit, and sample orbits are shown in Fig. \ref{fig:jeffery}c as closed paths on a sphere of the unit radius with the Bretherton parameter $B=0.9$. Note that when $|B|$ exceeds 1, the particle will align along a preferred orientation and no longer exhibit periodic motion.

\subsection{Hydrodynamic Symmetry}

Because the shape only affects the tensors in Eq. \eqref{eq:J03}, we can define the {\it hydrodynamic} shape by using the symmetry of the resistance tensors, rather than relying on the actual geometrical shape.

The symmetry arguments on the resistance tensors were introduced by Brenner \cite{brenner1964stokes2, brenner1964stokes3}, who followed the idea of hydrokinetic symmetry and argued for its connection with the kinetic energy of an ideal irrotational fluid in the nineteenth century \cite{kirchhoff1869, thomson1871, lamb}.

We consider a transformation of the body-fixed frame $\{\hat{\bm{e}}_i\}$ into another body-fixed frame $\{\hat{\bm{e}}'_i\}$. Let us denote the translation by ${\bf A}=(a_{ij})\in \textrm{O}(3)$,
which includes rotation and reflection. Now we write the matrix representations of an $r$th-rank tensor ${\bf X}$ for the body-fixed frames $\{\hat{\bm{e}}_i\}$ and $\{\hat{\bm{e}}'_i\}$ as $X_{i_i i_2, \cdots, i_r}$ and $X'_{i_i i_2, \cdots, i_r}$, respectively, and we obtain the following relation between the two representations:
\begin{equation}
X'_{i_1 i_2 \cdots i_r}=a_{i_1j_1} a_{i_2j_2} \cdots a_{i_rj_r} X_{j_i j_2 \cdots j_r}
\label{eq:J05a}.
\end{equation}
Similarly, we have the following relation between the matrix representations of an $r$th-rank pseudotensor ${\bf Y}$ under the transformation of the body-fixed frame:  
\begin{equation}
Y'_{i_1 i_2 \cdots i_r}= \sigma a_{i_1j_1} a_{i_2j_2} \cdots a_{i_rj_r} Y_{j_i j_2 \cdots j_r}
\label{eq:J05b},
\end{equation}
where we have introduced the determinant of the matrix ${\bl \textrm{det}{\bf A}}= \sigma=\pm1$. 

If the shape of an object is unchanged under a transformation, we usually argue that the object possesses the {\it geometrical} symmetry associated with the transformation matrix. 
Alternatively, we now introduce the {\it hydrodynamic} symmetry based on the invariance of the resistance tensors. In other words, if all the resistance tensors inside $\mathcal{K}$ and $\mathcal{G}$ are invariant under a change of the body-fixed frame denoted by ${\bf A}=(a_{ij})\in \textrm{O}(3)$, then the object is considered to possess hydrodynamic symmetry associated with ${\bf A}$. A transformation ${\bf A}$  induces geometrical symmetry and hydrodynamic symmetry for an object, but these two symmetries do not necessarily coincide. 

For example, if we take ${\bf A}$ as a rotational matrix around an axis by an arbitrary angle, the geometrical symmetry associated with this transformation is usually called an axisymmetric object, but more precisely this is a body of revolution. In contrast, the hydrodynamic symmetry associated with the same ${\bf A}$ is known as {\it helicoidal} symmetry \cite{brenner1964stokes2, brenner1964stokes3}, which includes a chiral object, being distinct from an achiral body of revolution. Further details are provided in Sec.\ \ref{sec:chiral}.

\section{Theoretical Extensions}
\label{sec:ext}

In this section, we review some theoretical extensions beyond the periodic Jeffery orbits of an axisymmetrical body.

\subsection{More General Shapes}
\label{sec:chiral}

The axisymmetric nature of the Jeffery equation \eqref{eq:J12c} enables us to project the angle dynamics onto a unit sphere on which the director vector $\bm{d}$ moves. This reduction is also available for the helicoidal symmetry introduced in the previous section. The helicoidal symmetry in the Stokes flow was first introduced by Brenner \cite{brenner1964stokes2, brenner1964stokes3}. 

As expected from its name, a simple helix has been often described as a helicoidal object. This simply allows hydrodynamical coupling between the translation and rotation in the direction of the symmetrical axis, and therefore a bacterial flagellum can corkscrew to propel the organism in viscous environments \cite{purcell1977life}. This representation has been validated for a rotating helical object such as a bacterial cell when averaged around the axis of rotation \cite{ishimoto2019n, ishimoto2019bacterial}. A simple helix, however, does not rigorously satisfy the definition of helicoidal symmetry unless the number of turns is an integer. Nonetheless,  numerical computations have confirmed that a simple helix is well approximated by a helicoidal object \cite{chen2011dynamical}. 

\begin{figure}[tbp]
\begin{center}
\begin{overpic}[width=7cm]{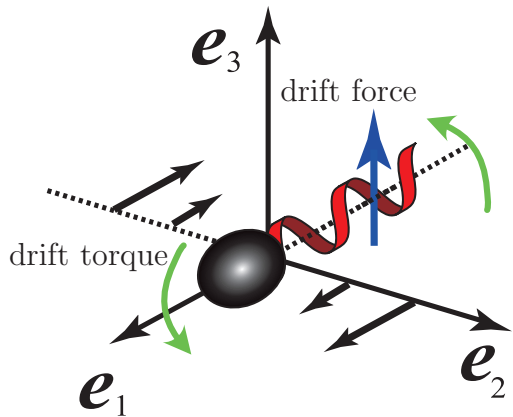}
\put(49,64){{\large drift force}}
\put(-2,33){{\large drift torque}}
\end{overpic}
\caption{ (Color online) Schematic of a helicoidal object under a simple shear. The chirality-induced forces induce a translational drift (shown with a blue arrow) and rotational drifts (shown with green arrows).}
\label{fig:helicoid}
\end{center}
\end{figure}

A chiral object generates a drift force when coupled to an external shear flow (Fig. \ref{fig:helicoid}). This force moves the chiral object away from the flow plane, and therefore one can use microfluidic devices to separate these particles depending on the handedness of their chirality \cite{makino2005migration, marcos2009separation, eichhorn2010microfluidic, aristov2013separation, ro2016chiral}.
When the chirality is not uniform along the axis of symmetry, such a chirality-induced drift then induces torque to orient the object towards the background vorticity vector.

With helicoidal symmetry, we may reduce the number of non-zero components in the resistance tensors as we did through Eqs. \eqref{eq:J10a}--\eqref{eq:J12a} for a body of revolution, which yields 13 non-vanishing constants of shape. Recently, Ishimoto \cite{ishimoto2020helicoidal} revisited the notion of Brenner's helicoidal symmetry and derived the angular dynamics of objects in this symmetry class, motivated by experiments of bacterial locomotion under a simple shear \cite{marcos2012bacterial} (further discussion of microswimmers in simple shear is provided in Sec.\ \ref{sec:bulk}). The final form of the dynamics \cite{ishimoto2020helicoidal} results in an additional term 
on the right-hand side of Eq. \eqref{eq:J12c}, given by
\begin{equation}
C\epsilon_{ijm}(\delta_{jk}-d_jd_k)E^\infty_{k\ell}d_\ell d_m
    \label{eq:E01}.
\end{equation}
This vector is perpendicular both to $\bm{d}$ and the term with $B$ in Eq. \eqref{eq:J12c}, and thus the trajectories of the director vector on a unit sphere move across Jeffery's orbits by this chirality-induced term when the object is under a simple shear (Fig. \ref{fig:helicoid}). Here, the shape-dependent constant $C$ is a new parameter representing the non-uniform distribution of the chiral coupling along the axis. Constructive calculations of this parameter were also made based on a detailed swimmer geometry \cite{mathijssen2019oscillatory}, and recent studies experimentally estimated the values of Ishimoto's shape parameter for bacterial cells \cite{jing2020chirality, zottl2022asymmetric} and swimming ciliates \cite{ohmura2021near}, and reported $C\sim O(10^{-2})$. This small number, however, generates the torque to align the cell towards the background vorticity vector in a simple shear, which we will see below.

The angle dynamics in a simple shear $\bm{u}^\infty=\gamma y\bm{e}_1$ are written for $\theta$ and $\phi$ in Fig. \ref{fig:jeffery}b as
\begin{equation}
\frac{d\theta}{dt}= h_1(\theta, \phi; \gamma, B, C), ~
\frac{d\phi}{dt}= h_2(\theta, \phi; \gamma, B, C) 
\label{eq:C01},
\end{equation}
where the functions $h_1$ and $h_2$ are respectively given by
\begin{eqnarray}
\gamma^{-1}h_1&=&\frac{B}{4}\sin 2\theta\sin 2\phi+\frac{C}{2}\sin\theta\cos2\phi \label{eq:C02a},\\
\gamma^{-1}h_2&=&-\frac{1}{2}(1-B\cos 2\phi)-\frac{C}{2}\cos\theta\sin 2\phi
\label{eq:C02b}.
\end{eqnarray}
These equations possess non-linear periodic orbits even for a non-zero $C$ if $B=0$. The integral curve is given by
\begin{equation}
C\sin2 \phi\sin^2\theta-2\cos\theta =\textrm{const}
    \label{eq:C02c}.
\end{equation}
If both $B$ and $C$ are non-zero, the degenerate periodic orbits become broken. In particular, when $B^2+C^2 < 1$, the poles $\theta=0, \pi$ are only the attractive/repulsive orientation, with the attracting pole being flipped by the change in the sign of $C$. The orientation vector eventually reaches the neighborhood of the stable pole and rotates with its attraction speed being slower than the exponential function. The change of the polar angle during one rotation near the poles is evaluated as $\Delta\theta\simeq K\theta^3$ with a non-zero constant $K=K(B, C)$.

\begin{figure}[htbp]
\begin{center}
\includegraphics[width=8.5cm]{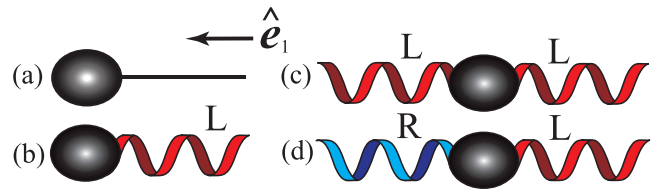}
\caption{ {\bl (Color online) Schematics of objects with helicoidal symmetry. (a) A spheroidal cell body with a slender rod is an example of a body of revolution. (b) A spheroidal cell body with a left-handed helical flagellum is well represented by an object with {\it helicoidal} symmetry. (c) A spheroidal cell body with two left-handed helical flagella. This possesses an additional $\pi$-rotational symmetry around the axis perpendicular to $\hat{\bm{e}}_1$ and is an example model of a {\it homochiral} object. (d) A spheroidal cell body with right-handed and left-handed helical flagella. This possesses an additional reflectional symmetry in a plane perpendicular to the axis of helicoidal symmetry, and an example model of a {\it hetrochiral} object.  Figure modified from Ishimoto \cite{ishimoto2020helicoidal}  under the creative commons license, http://creativecommons.org/licenses/by/4.0.} {\rd (copyright 2020 by The Author(s)) }}
\label{fig:bactmor}
\end{center}
\end{figure}

By definition, every body of revolution is a helicoidal object, and thus the helicoidal symmetries belong to a genuinely larger class of symmetry, including the body of revolution {\bl (Fig. \ref{fig:bactmor}a,b)}. Ishimoto \cite{ishimoto2020helicoidal} also found that the helicoidal symmetry contains two sub-classes of symmetry, called {\it homochiral} and {\it heterochiral} objects. A homochiral object possesses an additional $\pi$-rotational symmetry around an axis perpendicular to the symmetry axis {\bl (Fig. \ref{fig:bactmor}c)}. This induces a  fore-aft symmetric chirality distribution, yielding zero chiral rotational drift but non-zero chiral translational drift. An example is a simple helix whose rotational dynamics are the same as the achiral Jeffery equation \cite{chen2011dynamical,rost2020effective}. Heterochiral objects instead possess an additional mirror symmetry to the axis of symmetry, and therefore the chiral-induced translational drifts cease for these objects, but they still exhibit non-zero chiral rotational drifts {\bl (Fig. \ref{fig:bactmor}d)}. An example may be a composition of the same two helices but with opposite chirality connected in series. Kramel et al. \cite{kramel2016preferential} considered the dynamics of such particles in turbulent flow to detect local swirling structures of the flow field.


The concept of helicoidal symmetry was extended to discrete rotational symmetry. Indeed, Brenner \cite{brenner1964stokes3} introduced the notion of helicoidal symmetry by the invariance of the resistance tensors under $\pi/2$-rotational symmetry. 
Ishimoto \cite{ishimoto2020jeffery} confirmed that the dynamics of any object with $n$-fold rotational symmetry $(n\geq 4)$ satisfy the same equations under a linear flow as those for a helicoidal object. As in the continuous case, the rotational dynamics are reduced to achiral Jeffery equations by an additional mirror symmetry \cite{fries2017angular}. A complete classification of the translational and rotational dynamics is provided using the Schoenflies notions in Ref. \cite{ishimoto2020jeffery}.

Another interesting class of symmetry in terms of {\bl hydrodynamic} symmetry is Lord Kelvin's isotropic helicoid \cite{thomson1871}. This hypothetical object possesses a triaxial helicoidal symmetry, and its director dynamics are the same as that of a simple sphere \cite{brenner1964stokes3, gustavsson2016preferential} but with coupling between the translational and rotational motions. Recently, Collins et al. \cite{collins2021lord} fabricated a 3D-printed model and found that the particle predicted by Kelvin actually exists, though the transition-rotation coupling is extremely small.

When an object has three-fold rotational symmetry, however, additional shape parameters are needed to describe the motion dynamics in a linear flow. The rotational dynamics are no longer projected onto the unit sphere but require three angle variables, which may cause chaotic behaviors depending on the shape parameters \cite{ishimoto2020jeffery}. Further less symmetric cases with two-fold rotational symmetry contain an important class of symmetry, that is, tri-axially ellipsoidal particles  \cite{jeffery1922motion, hinch1979rotation}. The angular dynamics of a tri-axially mirror-symmetric particle are described by three shape parameters reflecting the length scales of each axis \cite{bretherton1962motion, brenner1964stokes3}, and an ellipsoidal particle can also exhibit chaotic rotational dynamics in a shear flow \cite{yarin1997chaotic, thorp2019motion}.

\subsection{Asymptotic Jeffery's Orbits}
\label{sec:asymp}

\begin{figure}[tbp]
\begin{center}
\begin{overpic}[width=8.5cm]{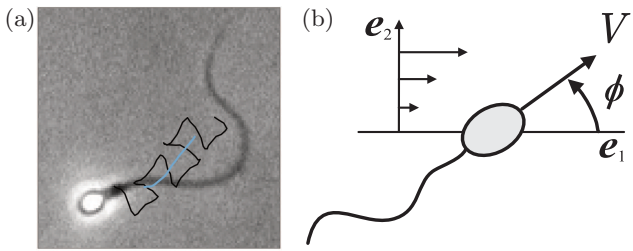}
\put(-3,37){{ (a)}}
\put(43,37){{ (b)}}
\end{overpic}
\caption{ (Color online) (a) Swimming trajectory of a human sperm head as an example of a yawing microswimmer traced from video microscopic images in Ishimoto et al. \cite{ishimoto2017coarse}. (b) Schematic of a model microswimmer under simple shear. Figures adapted from Walker et al. \cite{walker2022effects} under the creative commons license, http://creativecommons.org/licenses/by/4.0. {\rd (copyright 2022 by The Author(s)) }}
\label{fig:yaw}
\end{center}
\end{figure}

In microswimming dynamics, swimming is usually achieved by non-reciprocal deformations described by the scallop theorem \cite{purcell1977life}. In viscous environments, swimming is typically very inefficient, and large portions of displacements are canceled even with rapid oscillation, as illustrated by the large yawing motion of sperm motility shown in Fig. \ref{fig:yaw}a \cite{ishimoto2017coarse}. 

When two separable timescales exist in the dynamics, one may be more interested in the slow dynamics after averaging out the fast timescale. Walker et al. \cite{walker2022effects} introduced the classical multiple-scale expansions to derive the asymptotic slow dynamics for rapidly yawing microswimmers in a shear flow. Here, we provide the essential analyses in Walker et al. \cite{walker2022effects} for planar swimming, focusing on a simple sinusoidal yawing motion (Fig. \ref{fig:yaw}b). We denote the position of the swimmer by $(x, y)$ and its angle from the $\bm{e}_1$-axis by $\phi$. The swimmer is assumed to be self-propelled in this orientation with a constant velocity $V$. Simple shear is induced in this plane by $\bm{u}^\infty=\gamma y\bm{e}_1$, and we write the observed yaw angle as a sum of the slow dynamics and fast sinusoidal oscillation in the form of $\phi(t)=\overline{\phi}(t)+A\sin(\omega t)$, where $A$ is the amplitude of the yaw. Assuming the passive part of the dynamics follows Jeffery's orbits \eqref{eq:J16}, we may write the swimmer dynamics as
\begin{eqnarray}
\frac{dx}{dt}&=&V\cos\phi+\gamma y,\label{eq:E11a} \\
\frac{dy}{dt}&=&V\sin\phi,\label{eq:E11b} \\
\frac{d\phi}{dt}&=&-\frac{\gamma}{2}(1-B\cos2\phi)+A\omega\cos(\omega t)
\label{eq:E11c}.
\end{eqnarray}

Following the classical procedures of multiple-scale analysis, we introduce the fast variable $T:=\omega t$ with $\omega \gg 1$ and the formal transformation $t\mapsto (t, T)$ and write $x=x(t, T), y=y(t, T)$, and $\phi=\phi(t, T)$. The angle evolution \eqref{eq:E11c} is then rewritten as
\begin{equation}
\frac{\partial \phi}{\partial t}+\omega\frac{\partial \phi}{\partial T}=-\frac{\gamma}{2}(1-B\cos2\phi)+A\omega\cos T
\label{eq:E12}.
\end{equation}
We then pose an asymptotic expansion for $\phi$ in terms of inverse series of $\omega$ in the form of $\phi=\phi_0+\omega^{-1}\phi_1+\cdots$. The leading-order problem for $\omega\gg 1$ becomes $\partial_T\phi{\bl _0=A}\cos T$; hence,
$\phi_0=\overline{\phi}_0(t)+A\sin T$, where $\overline{\phi}_0$ denotes the slow dynamics at the leading order. The slow dynamics are determined by the solvability condition that corresponds to the $2\pi$-periodicity in $T$. We denote the $T$ average by $\langle ~\rangle$ to obtain 
\begin{equation}
\frac{d\overline{\phi}{\bl _0}}{dt}=-\frac{\gamma}{2}(1-B\langle \cos 2\phi_0\rangle)
\label{eq:E13},
\end{equation}
if the next-order dynamics are averaged. Similar analyses for the translational dynamics enable us to derive the asymptotic behaviors of the swimmer $(\overline{x}_0,\overline{y}_0,\overline{\phi}_0)$ in the form 
\begin{eqnarray}
\frac{d\overline{x}_0}{dt}&=&V_e\cos\phi+\gamma \overline{y}_0,\label{eq:E14a} \\
\frac{d\overline{y}_0}{dt}&=&V_e\sin\phi,\label{eq:E14b} \\
\frac{d\overline{\phi}_0}{dt}&=&-\frac{\gamma}{2}(1-B_e\cos2\overline{\phi}_0)
\label{eq:E14c},
\end{eqnarray}
where $V_e=J_0(A)V$ and $B_e=J_0(2A)B$ are the effective swimming speed and the effective Bretherton parameter for the asymptotic slow dynamics, respectively, and $J_0(A)=\langle \cos(A\sin T)\rangle$ is the zeroth-order Bessel function of the first kind. Comparing Eqs. \eqref{eq:E11a}--\eqref{eq:E11c} with \eqref{eq:E14a}--\eqref{eq:E14c} shows that the slow dynamics satisfy the same Jeffery orbits with a renormalized speed and shape parameter. With this sinusoidal yawing, the effective shape of the particle approaches a sphere as the function $J_0(A)$ gradually damps to zero with oscillation in the increase of $A$. Further generalizations of the yawing function are made in Ref. \cite{walker2022effects}.

The simple example discussed above assumed that the angular dynamics follow Jeffery's orbits in the absence of the yawing motion. Gaffney et al. \cite{gaffney2022canonical} relaxed this assumption and generalized the asymptotic Jeffery's orbits to a wide class of shape-changing swimmers. They considered a drift-free swimmer with a periodic shape gait in the plane of a shear flow.  Let the shear rate $\gamma$ and frequency of the shape deformation $\omega$ satisfy $\omega \gg \gamma$, and then the average rotational motion of the swimmer is given by a Jeffery's orbit with a relative asymptotic error of $O(\gamma/\omega)$. Here, the multiple-scale expansions are again performed in the resistance tensors $\mathcal{K}_{ab}$ and $\mathcal{G}_{ajk}$, which may be treated as functions of only the fast-time scale when represented in the body-fixed frame. The asymptotic angular dynamics are
\begin{eqnarray}
\frac{d\overline{\phi}_0}{dt}=-\frac{\gamma_e}{2}\left[ 1-B_e\cos(2\overline{\phi}_0+2\phi_c)\right]
    \label{eq:E15},
\end{eqnarray}
where $\gamma_e$ is the effective shear strength, $B_e$ is the effective Bretherton parameter, and $\theta_c$ represents the phase shift. 

The asymptotic Jeffery's orbits found in the slow dynamics are essential features of the microswimmer dynamics. Indeed, numerical simulations of a sperm cell and flagellate in a shear flow well reproduced Jeffery's orbits on average \cite{walker2018response, walker2022effects}. These results include a rigid body with an asymmetric shape, and thus the dynamics trace the Jeffery orbits. With a boomerang shape with two rods, $B_e$ may exceed 1 so that the swimmer orientation can be aligned rather than exhibit rotation\cite{roggeveen2022motion}.

Another type of timescale separation can arise when the swimmer, such as a bacterium cell, is rapidly spinning along its elongated direction. Dalwadi et al. \cite{dalwadi2023part1, dalwadi2023part2} considered a helicoidal squirmer with an arbitrary constant surface slip swimming under a background shear flow, $\bm{u}^\infty=\gamma y\bm{e}_1$. We take the axis of the helicoidal symmetry as $\bm{d}=\hat{\bm{e}}_1$, as in Fig. \ref{fig:coord}, and denote the self-propelled linear velocity by $\bm{V}$ and the angular velocity by $\bmOmega^{\textrm{prop}}=\itOmega_\|\hat{\bm{e}}_1+\itOmega_\perp\hat{\bm{e}}_2$. After introducing the appropriate Euler angles $(\theta, \psi, \phi)$, the angular dynamics of the swimmer are given by 
\begin{eqnarray}
\frac{d\theta}{dt}&=& \itOmega_\perp\cos\psi+h_1(\theta, \phi; \gamma, B, C)  \label{eq:E16a},\\
\frac{d\phi}{dt}&=& \itOmega_\perp\csc\theta\sin\psi+h_2(\theta, \phi; \gamma, B, C) \label{eq:E16b},\\
\frac{d\psi}{dt}&=& \itOmega_\|-\itOmega_\perp\cot\theta\sin\psi+h_3(\theta, \phi; \gamma, B, C, D) \label{eq:E16c}.
\end{eqnarray}
Here, the functions $h_1, h_2$, and $h_3$ represent the shear-induced dynamics in each angle variable; $h_1$ and $h_2$ are provided in Eqs. \eqref{eq:C02a} and \eqref{eq:C02b}; and $h_3$ is given by
\begin{equation}
\gamma^{-1}h_3=-\frac{B}{2}\cos\theta\cos 2\phi+\frac{C}{2}\cos^2\theta \sin2\phi+\frac{D}{2}\sin^2\theta \sin2\phi
\label{eq:E17}.
\end{equation}
$\itOmega_\|$ affects only the $\psi$ angle, the spinning angle, and when $\itOmega_\perp=0$, the {\bl equations} are closed in the angles $\theta$ and $\phi$, leading to the dynamics of a helicoidal object {\bl , Eqs.\eqref{eq:C01}} . Indeed, the functions $h_1$ and $h_2$ contain the shape parameters $B$ and $C$, whereas the new chirality-induced shape parameter $D$ should appear in the function $h_3$.

Let us introduce the ratio of the two components of the angular velocity $p$ as $\itOmega_\perp=p \it\Omega_\|$. When $p=0$, the self-propulsion only induces spinning along the symmetry axis, as often assumed for a model bacterium \cite{purcell1977life, lauga2006swimming}. In contrast, when $p$ increases as $p\rightarrow \infty$, the swimmer trajectory approaches a circle in a plane, as observed in marine sperm cells  \cite{riedel2005self, shiba2008ca2+, ishimoto2015fluid}.

In the absence of a background flow, the trajectory of such a swimmer traces a single helix, and this behavior is expressed by the helix theorem\cite{shapere1989geometry}. These helical trajectories are reported in many biological microswimmers, such as microalgae \cite{bearon2013helical, de2020motility, cortese2021control} and sperm cells \cite{friedrich2007chemotaxis}, as well as artificial swimmers \cite{yamamoto2017chirality}. Direct integration of Eqs. \eqref{eq:E16a}--\eqref{eq:E16c} after setting $h_1=h_2=h_3=0$ allows one to derive the radius and pitch of the helical trajectories as 
\begin{equation}
r_{\textrm{helix}}=\frac{\itOmega_\perp}{\itOmega_\|^2+\itOmega_\perp^2}V~,~~b_{\textrm{pitch}}=\frac{2\pi \itOmega_\|}{\itOmega_\|^2+\itOmega_\perp^2}V
    \label{eq:E15}.
\end{equation}
{\bl Here we have taken the swimming direction as $\hat{\bm{e}}_1$ so that the swimming speed $V$ is obtained from $\bm{V}=V\hat{\bm{e}}_1$.}

Now we consider the timescale separation of self-induced spinning and shear-induced rotation, assuming $\itOmega_\|=O(\itOmega_\perp)\gg \gamma$. The fast timescale is denoted as $T={\bl (}\itOmega_\|^2+\itOmega_\perp^2)^{1/2} t$, where $t$ denotes the slow timescale.  As in the rapid yawing case, we expand the angle variables by the inverse series of $\lambda:=(1+p^2){\bl ^{1/2}}\itOmega_\|/\gamma\gg 1$ and use similar notation to perform multiple-scale analysis, yielding the slow dynamics with leading-order correction as
\begin{eqnarray}
\frac{d\overline{\theta}_0}{dt}&=& h_1(\overline{\theta}_0, \overline{\phi}_0; \gamma, \hat{B}, \hat{C})  \label{eq:E17a},\\
\frac{d\overline{\phi}_0}{dt}&=& h_{\bl 2}(\overline{\theta}_0, \overline{\phi}_0; \gamma, \hat{B}, \hat{C}) \label{eq:E17b}, \\
\frac{d\overline{\psi}_0}{dt}&=& h_{\bl 3}(\overline{\theta}_0, \overline{\phi}_0; \gamma, \hat{B}, \hat{C}, \hat{D}). ~~~~~~~~\label{eq:E17c}
\end{eqnarray}
This set of equations \eqref{eq:E17a}--\eqref{eq:E17c} completely reproduce the original angular dynamics \eqref{eq:E16a}--\eqref{eq:E16c} {\bl without the self-propelled rotation terms,} after introducing the effective shape parameters as
\begin{equation}
\hat{B}= \frac{(2-p^2)B}{2(1+p^2)},~
\hat{C}= \frac{C+p^2D}{(1+p^2)^{3/2}},~ 
\hat{D}= \frac{3p^2 C+(2-p^2)D}{2(1+p^2)^{3/2}} \label{eq:E18}.
\end{equation}
This result validates the descriptions of the helicoidal object \eqref{eq:C01} by introducing the effective shape parameters $\hat{B}$ and $\hat{C}$, depending on the rapid spinning behaviors.
 Indeed, the asymptotic dynamics recover the chiral Jeffery equations \eqref{eq:C01} and \eqref{eq:C02b} as $p\rightarrow 0$. In the opposite limit of $p\rightarrow \infty$, the trajectory loses its chirality effects as it traces Jeffery's orbits but with  $\hat{B}\rightarrow -B/2$. Similar asymptotic behaviors are also analyzed for the translational dynamics \cite{dalwadi2023part1, dalwadi2023part2}.


\subsection{Breakdown of Degenerate Orbits: Inertia, Rheology, and Other Factors}

The breakdown of the degeneracy in Jeffery's orbits has been studied from various perspectives, which were reviewed by Leal\cite{leal1980particle} in the early days of  research in this topic. 
The effects of small amounts of fluid {\bl inertia} have been investigated by perturbation theory with a small Reynolds number, $Re$, and it has been found that the inertia induces a drift towards rotation in a flow-gradient plane ($\bm{e}_1\bm{e}_2$ plane in Fig. \ref{fig:jeffery}a) as a stable limit cycle, whereas the poles become repellers \cite{subramanian2005inertial}. In addition to the fluid inertia, analysis of the effects of particle inertia, with its magnitude being characterized by the Stokes number, $St$, found that the in-plane rotation can be stabilized or destabilized to flip the stability between the in-plane rotation and poles, while these depend on the amount of inertia and shape of the spheroid  \cite{subramanian2006inertial, lundell2010heavy, einarsson2015effect, einarsson2015rotation}. It was also found that with an increase of the inertial effect, the rotational dynamics can exhibit a transition to chaos \cite{lundell2011effect, rosen2017chaotic}.

Degeneracy in the Jeffery orbits can also be broken by non-Newtonian rheology \cite{saffman1956motion, leal1975slow}. A classical analysis by Leal \cite{leal1975slow} showed a similar drift effect on a fiber in a viscoelastic (second-order) fluid, leading to the orientation of particles away from in-plane rotation. Several theoretical and computational analyses on the effect of the viscoelastic model have been performed with different shapes, rheological models, and parameters \cite{brunn1977slow, leal1979motion, brunn1980motion, d2014bistability, abtahi2019jeffery, ferec2021rigid}. Interested readers are referred to a recent review paper by Li et al. \cite{li2022some} and references therein.

The breakdown of the degenerate Jeffery's orbits also occurs by particle deformability and has been intensively studied for a droplet and a capsule \cite{zhao2011dynamics, omori2012reorientation, cordasco2013orbital, zhang2019dynamics},
as well as flexible fibers \cite{tornberg2004simulating, nguyen2014hydrodynamics, du2019dynamics, zuk2021universal, slowicka2020flexible}. The effects of the slip-boundary conditions have been numerically investigated by Zhang et al.\cite{zhang2015anisotropic}, who found that the orientation dynamics follow the Jeffery orbit under a certain symmetry, whereas recent simulations \cite{gravelle2021violations, kamal2021alignment} report the alignment of a particle in a flow plane.


\section{Microswimmers in Flows}
\label{sec:flow}

In this section, we review theoretical models and experimental observations in the studies of the motion of a single microswimmer in simple to complex background flow environments.

\subsection{Simple Shear}
\label{sec:bulk}

We start with a self-propelled particle in a simple shear flow, and many parts of this analysis are shown in the previous sections. Not only a rigid body, but shape-changing bodies such as sperm cells and {\it Leishmania} have been computationally studied and found to periodically rotate along Jeffery orbits but with a drift due to the background shear \cite{walker2018response, kumar2019effect, walker2022effects}. When a swimmer possesses chirality, such as bacterial flagella, the additional shear-induced drift force and torque orient the swimmer towards the background vorticity vector, which is perpendicular to the flow plane (Sec. \ref{sec:chiral}). Bacterial swimmers, therefore, move with non-zero perpendicular velocity, and this biased locomotion is known as bacterial bulk rheotaxis \cite{marcos2012bacterial, jing2020chirality, zhang2021base, ronteix2022rheotaxis, zottl2022asymmetric}.

As introduced in Sec.\ \ref{sec:asymp}, a squirmer swimmer with a time-constant surface profile in general exhibits a helical path. Bearon \cite{bearon2013helical} studied a similar mathematical model with helical trajectories in the presence of gravitational torque and a simple shear flow to understand the dynamics of swimming microalgae such as {\it Chlamydomonas} and {\it Heterosigma akashiwo} under a turbulent flow. The study of spherical objects found a novel stable equilibrium in a vertical shear flow, suggesting a robust upward swimming due to the helical path.

\begin{figure}[tbp]
\begin{center}
\begin{overpic}[width=8.5cm]{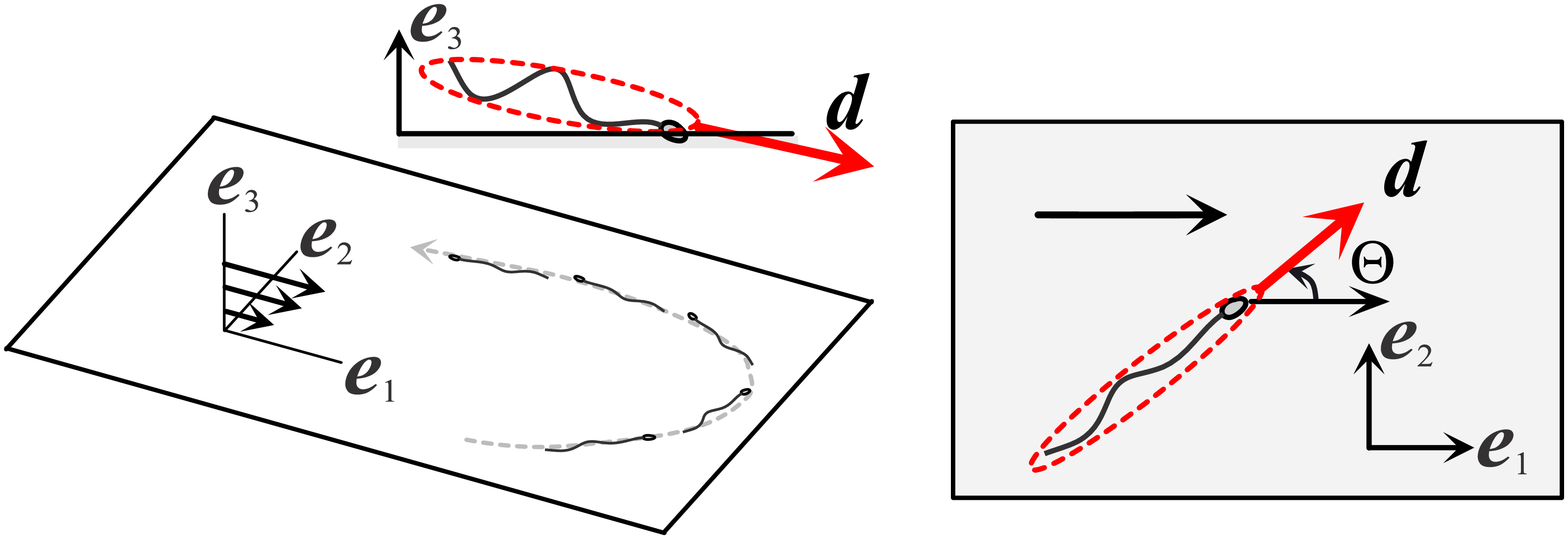}
\put(3,31){{\large (a)}}
\put(55,31){{\large (b)}}
\put(61,23){{flow direction}}
\end{overpic}
\caption{ (Color online) Schematic of a rheotactic microswimmer under a simple shear flow near a flat wall. (a) Schematic of the swimmer behavior and flow configuration. (inset) Side view of the swimmer facing toward the wall with its director being denoted by a unit vector $\bm{d}$. (b) Projection onto the wall plane ($\bm{e}_1\bm{e}_2$-plane). The in-plane angle $\Theta$ is taken from the $\bm{e}_1$-axis (flow direction). By this weather-vane mechanism, a microswimmer with a wall-attacking orientation turns its swimming direction upstream.}
\label{fig:rheo_theor}
\end{center}
\end{figure}

A simple shear flow is also ubiquitous in the vicinity of an external boundary such as a flat wall (Fig. \ref{fig:rheo_theor}). This geometrical confinement is almost unavoidable for laboratory experiments and also plays significant roles in microbial habitats   \cite{conrad2018confined} and reproduction processes, in particular, for internal fertilizers\cite{gaffney2011mammalian, nosrati2017microfluidics}. Indeed, it is well known that these cells accumulate near boundaries due to contact force and hydrodynamic interactions, as intensively studied in bacteria \cite{berke2008hydrodynamic, shum2010modelling, bianchi2017holographic}, sperm \cite{smith2009human, elgeti2010hydrodynamics, ishimoto2014study}, and squirmer models \cite{ishimoto2013squirmer, uspal2015rheotaxis}.  
Gravity also contributes to the sedimentation of artificial microswimmers and confines them in the vicinity of a bottom surface.

\begin{figure*}[htbp]
\begin{center}
\begin{overpic}[width=18.5cm]{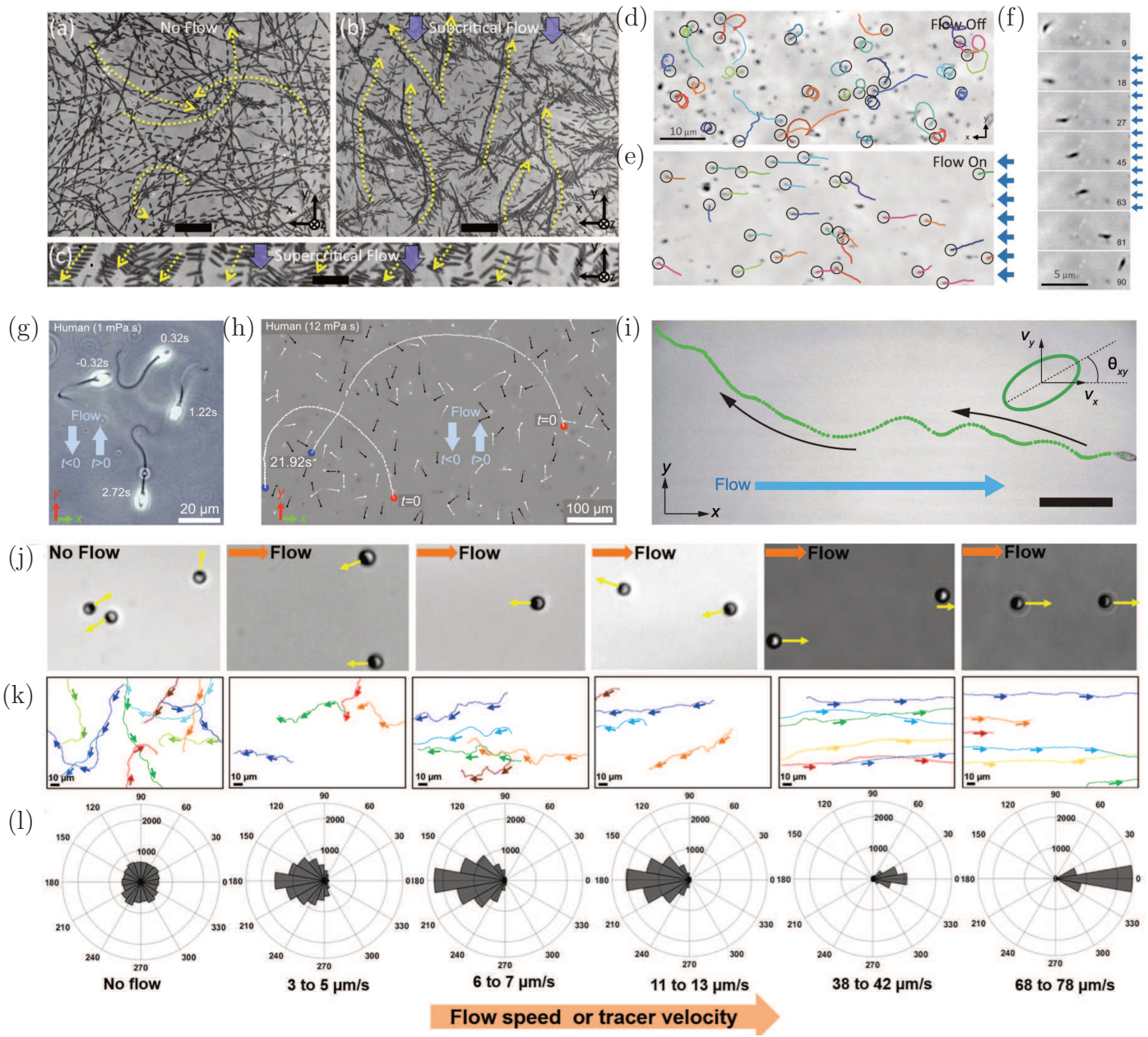}
\put(52,88){{\large (d)}}
\put(52,76){{\large (e)}}
\put(85,88){{\large (f)}}
\put(-0.5,61){{\large (g)}}
\put(18,61){{\large (h)}}
\put(52,61){{\large (i)}}
\put(-0.5,41){{\large (j)}}
\put(-1,29){{\large (k)}}
\put(-0.5,18){{\large (l)}}
\end{overpic}
\caption{ (Color online) (a--c) Sample trajectories of {\it E. coli} bacteria swimming in the vicinity of a flat wall. Superimposed snapshots are shown to illustrate the trajectories of individual cells, which are shown by dotted arrows. (a) Bacterial cells swim in circles due to hydrodynamic interactions between the cell and wall in the absence of a background flow. (b) Flow induction (from the top of the figure, shown in blue arrows) triggers bacterial rheotactic behavior, resulting in upstream migration. (c) At a higher shear rate, cells are swept away by the flow. Figure adapted from Kaya et al. \cite{kaya2012direct}  with permission (copyright 2012 by Elsevier). (d--f) Sample trajectories of {\it M. pneumoniae} cells under a gliding motility on a glass substrate. Cell trajectories are in the (d) absence and (e) presence of a flow. The blue arrows indicate the flow direction from right to left. The positions at the initial time are denoted by colored circles.  (f) Time-lapsed images of a single cell exhibiting rheotactic behavior. Figure adapted from Nakane et al. \cite{nakane2022cell} with permission under the creative commons license, http://creativecommons.org/licenses/by/4.0 {\rd (copyright 2022 by The Author(s)) }.  (g,h) Experimental observations of rheotactic human sperm cells. (g) A sudden change of swimming orientation by a reversal of the flow direction is shown by superimposed images of a swimming cell at different times. The flow is reversed at $t=0$. (h) Similar experimental observations, but with different viscosity of the swimming medium, show slower responses of the cells. Figure adapted from Kantsler et al.\cite{kantsler2014rheotaxis} with permission under the creative commons license, http://creativecommons.org/licenses/by/4.0 {\rd (copyright 2014 by The Author(s)) }. (i) Sample trajectory of a rheotactic {\it T. pyriformis} cell with the background flow moving from left to right, whereas the cell swims upwards along the bottom wall. The scale bar is 200 $\mu$m. Figure adapted from Ohmura et al. \cite{ohmura2021near} with permission under the creative commons license, http://creativecommons.org/licenses/by/4.0 {\rd (copyright 2021 by The Author(s)) }. (j--l) Various behaviors of a catalytic Janus particle near a flat wall under a flow with different strengths of shear rate. The flow direction is indicated by the orange arrow at the bottom (from left to right). Each row indicates (j) snapshots of optical images, (k) sample trajectories of Janus particles, and (l) distributions of the directions of motion. At the intermediate shear strength, particles can migrate upstream, while the high shear flow sweeps the particles away from the near-wall region. Figure adapted with permission from Sharan et al.\cite{sharan2022upstream} (copyright 2022 by the American Chemical Society).}
\label{fig:rheo_exp}
\end{center}
\end{figure*}

A notable phenomenon of a microswimmer near a wall in the presence of a background shear is the swimming behavior against the flow, which is known as (surface) {\it rheotaxis}. It has been reported that many microswimmers exhibit this upstream migration ranging from biological cells to artificial active colloids and that these behaviors are often explained by simple mechanical models (Fig. \ref{fig:rheo_exp}). 
Theoretical building blocks for the various hydrodynamic and mechanical interactions are examined in Mathijssen et al. \cite{mathijssen2019oscillatory}. Here, however, we only sketch out how these rheotactic behaviors emerge from the shear-induced hydrodynamic interactions, using a simple model by Kantsler et al.\cite{kantsler2014rheotaxis}. 

We consider an object whose rotational dynamics obey the Jeffery equation.  
The presence of a rigid flat wall near Jeffery orbits yields only a slight alternation on the rotational dynamics for a prolate spheroid, although it is more significant for an oblate spheroid \cite{pozrikidis2005orbiting}. We let $\bm{d}$ be the unit vector for the swimmer and consider a simple shear $\bm{u}^\infty=\gamma z\bm{e}_x$ with $z=0$ being the flat wall. We assume that the attack angle towards the flat wall $\cos^{-1}(\bm{d}\cdot\bm{e}_3)$ is kept constant in time based on experimental observations, and it reflects steric, hydrodynamic, and other mechanical interactions. The Jeffery equation is then reduced to 
\begin{eqnarray}
\gamma^{-1}\dot{d}_1&=&\frac{B}{2}d_3(1-2d_1^2)+d_3+\lambda d_1 
    \label{eq:R01a} \\
\gamma^{-1}\dot{d}_2&=&-Bd_1d_2d_3+\lambda d_2
\label{eq:R01b},
\end{eqnarray}
where $\lambda$ is a constant determined by the constraint on the attack angle. Substituting the value of $\lambda$ further simplifies the angle dynamics to 
\begin{equation}
\dot{d}_1=\gamma\frac{(1+B)}{2}\frac{d_2^2d_3}{1-d_3^2},~
\dot{d}_2=-\gamma\frac{(1+B)}{2}\frac{d_1d_2d_3}{1-d_3^2}
\label{eq:R02}.
\end{equation}
Introducing an in-plane angle $\Theta$ as $d_1=(1-d_2^3)^{1/2}\cos\Theta$ and $d_2=(1-d_2^3)^{1/2}\sin\Theta$ (see also Fig. \ref{fig:rheo_theor}b), the in-plane dynamics \eqref{eq:R02} are thus written as
\begin{eqnarray}
    \frac{d\Theta}{dt}=-\alpha \sin\Theta
    \label{eq:R03},
\end{eqnarray}
where $\alpha=(\gamma/2)(1+B)d_3$. For a typical object, $|B|<1$ is satisfied, and thus the sign of $\alpha$ is determined by the sign of $d_3$. When the swimmer's head is facing towards the wall boundary as shown in  Fig. \ref{fig:rheo_theor}a, according to Eq. \eqref{eq:R03}, the downstream direction ($\Theta=0$) becomes unstable and the in-plane orientation eventually evolves to the upstream direction ($\Theta=\pi$). 
Note also that the rheotactic response is faster for an elongated object by an increase of $B$, but even a spherical object ($B=0$) can exhibit the same rheotactic response. The hydrodynamic interactions that mechanically cause the rheotactic behavior of microswimmers are sometimes referred to as weather-vane mechanisms. In the above arguments, note that we have neglected hydrodynamic wall interactions, steric interactions, and chiral shape effects.  

Surface rheotaxis has been studied in swimming
 bacteria \cite{hill2007hydrodynamic, kaya2012direct, mathijssen2019oscillatory}. Kaya et al. \cite{kaya2012direct} experimentally demonstrated surface rheotaxis in {\it E. coli} (Fig. \ref{fig:rheo_exp}a--c). The bacterial cells swim in circles near a wall boundary via the hydrodynamic torque generated by their spinning flagella \cite{lauga2006swimming} in the absence of a background flow (Fig. \ref{fig:rheo_exp}a). When external flow is induced, the bacterial cells turn upstream and exhibit surface rheotaxis at the moderate shear rate (Fig. \ref{fig:rheo_exp}b). At a higher shear rate, however, bacterial cells are swept away by the external flow (Fig. \ref{fig:rheo_exp}c). Many bacteria can migrate on the substrate without using flagella but with other short and thin appendages called type-IV pili.
 The upstream migration by pili-mediated motile bacteria has been reported in various species such as {\it Pseudomonas aeruginosa}\cite{shen2012flow}, {\it Acidovorax citrulli}\cite{meng2005upstream}, and 
{\it Xylella fastidiosa}\cite{bahar2010assessing}. Another type of gliding motility on the substrate is known in {\it Mycoplasma} bacteria, and they also exhibit upstream migration \cite{rosengarten1988rheotactic, nakane2022cell}. Fig. \ref{fig:rheo_exp}d--f show the experimental observation of {\it Mycoplasma pneumoniae} by Nakane et al. \cite{nakane2022cell}. During the gliding motility, cells adhere to the substrate and exhibit membrane protrusion in the leading ``edge," forming a wall-attack configuration as in Fig. \ref{fig:rheo_theor}a. Sample trajectories and time-lapsed images of {\it M. pneumoniae} cells in Fig. \ref{fig:rheo_exp}e and f show cell migration towards the right, facing upstream. This bacterial rheotaxis is also biologically relevant to understanding the invasion of infectious bacteria in medical devices and within organs.

Sperm rheotactic behavior was first reported in the 1870s \cite{lott1872} for mammalian sperm, and its mechanisms were debated for more than a century. Bretherton and Rothschild \cite{bretherton1961rheotaxis} examined sperm rheotaxis, but they could not explain the rheotactic behavior of live cells. More recently, however, hydrodynamic interaction has successfully explained mammalian sperm rheotactic behaviors \cite{miki2013rheotaxis, kantsler2014rheotaxis, ishimoto2015fluid, tung2015emergence, omori2016upward}. Fig. \ref{fig:rheo_exp}g and h reproduce images from Kantsler et al. \cite{kantsler2014rheotaxis} that show superimposed images and sample trajectories of rheotactic human sperm cells in different media. Experiments with calcium imaging analysis also support the mechanical origin of this phenomenon rather than the hypothesis that cells sense the flow and actively modulate their motility \cite{zhang2016human}.  The recent development of microfluidics device further advances
quantitative analysis and sperm sorting based on sperm rheotaxis
\cite{rappa2018quantitative, zaferani2018rheotaxis, sarbandi2021rheotaxis, hyakutake2021experimental, sharma2022selection}. Bukatin et al. \cite{bukatin2015bimodal} reported two modes of rheotactic turning, suggesting the roles of flagellar elasticity  \cite{bukatin2015bimodal}.
These elastic properties are known to be important for prompting  collective cell behavior \cite{taketoshi2020elasto, schoeller2020collective}. Recently, collective sperm rheotaxis has been reported, such as in chicken sperm  \cite{el2022new} and 
bull sperm \cite{phuyal2022biological}, suggesting swimming abilities are enhanced when individual sperm are part of a group.

Other biological cells are known to exhibit surface rheotaxis. Ohmura et al. \cite{ohmura2021near} recently found that the swimming ciliate {\it Tetrahymena pyriformis} exhibits surface rheotaxis, and their experimental and numerical observations suggest the kinesthetic sensing system of cilia. Fig. \ref{fig:rheo_exp}i shows a sample trajectory of {\it T. pyriformis} under a simple shear in the vicinity of a bottom wall, as reported in Ohmura et al. \cite{ohmura2021near}. Attachments of the cilia to the bottom substrate alter the ciliary beating, resulting in asymmetric propulsion that maintains the wall-attacking orientation of the cell. Rheotactic behaviors are also reported in a crawling ciliate, {\it Uronychia setigera} \cite{ricci1999rheotaxis}, that employs its cilia for locomotion on the substrate. Upstream swimming is also known in {\it C. elegans} \cite{yuan2015propensity, ge2019profile}. In this review, we have highlighted the physical and hydrodynamic mechanisms of the rheotaxis; however, note that biological cells usually possess several mechanosensing channels and can respond to mechanical cues. Indeed,  {\it Dictyostelium} rheotaxis is dependent on sensory response \cite{lima2014role}. The mechanisms underlying experimental observation should be carefully analyzed.

Surface rheotaxis is also realized in synthetic active colloids and microrobots, such as light-activated colloidal particles\cite{palacci2015artificial}, half-coated catalytic Janus particles \cite{uspal2015rheotaxis, katuri2018cross, sharan2022upstream}, self-propelled Au-Pt nanorods \cite{ren2017rheotaxis, baker2019fight, brosseau2019relating}, and active droplets \cite{dwivedi2021rheotaxis, dey2022oscillatory}. Fig. \ref{fig:rheo_exp}j--l, adapted from Sharan et al. \cite{sharan2022upstream}, show the motions of catalytic Janus particles under a background flow near a plane boundary. The panels represent experimental optical images (Fig. \ref{fig:rheo_exp}j), sample trajectories (Fig. \ref{fig:rheo_exp}k), and the distribution of the motion direction (Fig. \ref{fig:rheo_exp}l) under various shear rate strengths. 
{\bl By considering the surface velocity slip, gravity, and surface repulsion, their numerical simulations of a squirmer model well explain the various rheotactic behaviors.} 
Theoretical investigations for the rheotactic response of microswimmers have been performed for some simple mathematical models, such as a squirmer  \cite{uspal2015rheotaxis, ishimoto2017guidance, ishimoto2017dynamics} and a three-sphere model  \cite{daddi2020tuning}. The rheotactic behavior is affected by the flow field that the swimmer itself creates because this field alters the near-surface behavior.


\subsection{Poiseuille Flow}

\begin{figure}[tbp]
\begin{center}
\begin{overpic}[width=8.5cm]{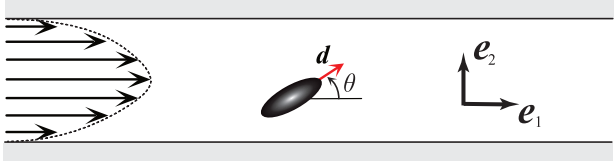}
\end{overpic}
\caption{ (Color online) Schematic of a microswimmer in planar Poiseuille flow.  
}
\label{fig:pois_theor}
\end{center}
\end{figure}

Next, we proceed to a microswimmer in Poiseuille flow with quadratic flow and a unidirectional velocity profile (Fig. \ref{fig:pois_theor}). We assume that the microswimmer is sufficiently small compared to the width of the chamber so that the background flow around the swimmer is well approximated by the linear flow. As shown in Fig. \ref{fig:pois_theor}, we consider a planar Poiseuille flow with a chamber width $h$ and focus on the swimmer movements in the $\bm{e}_1\bm{e}_2$ plane, where the flow and chamber direction are taken as the $\bm{e}_1$-axis. The background flow field is then written as $\bm{u}^\infty=\gamma(h^2-y^2)\bm{e}_1$, with $\gamma$ representing the strength of the flow. Let the model microswimmer move to the orientation vector $\bm{d}=(\cos\theta, \sin\theta)$ with a constant speed $V$ and assume that the orientation dynamics follow the Jeffery equation with the local linear flow. The evolutions of the position and the orientation of the swimmer are then written as
\begin{eqnarray}
    \dot{x}&=&V\cos\theta+\gamma(h^2-y^2) \label{eq:P01a}, \\
    \dot{y}&=&V\sin\theta \label{eq:P01b}, \\
    \dot{\theta}&=&\gamma y(1-B\cos2\theta)
    \label{eq:P01c},
\end{eqnarray}
where $B$ is the Bretherton parameter. We hereafter set $h=1$.

\begin{figure}[htbp]
\begin{center}
\begin{overpic}[width=4cm]{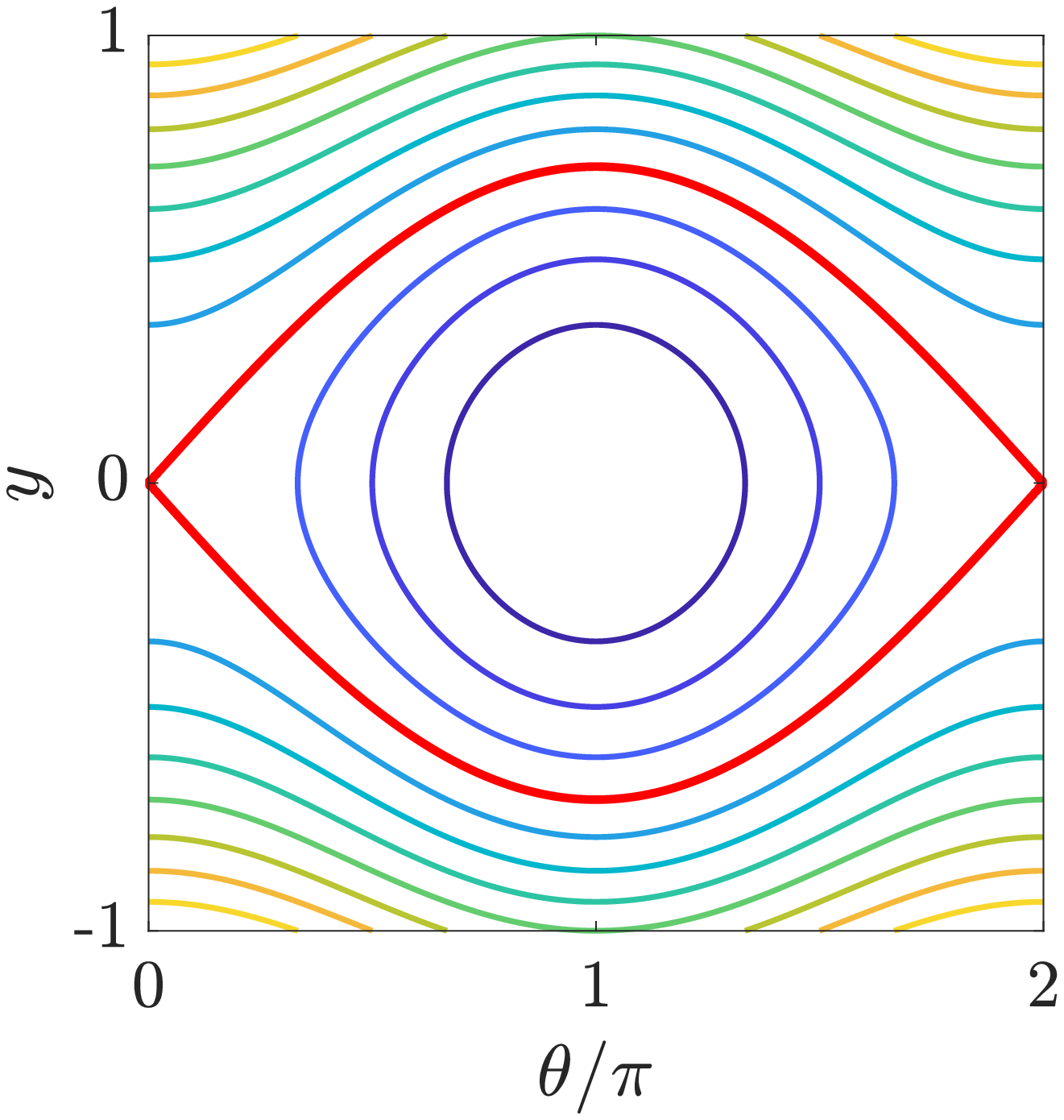}
\put(-6,95){{(a)}}
\end{overpic}
\hspace{0.5cm}
\begin{overpic}[width=4cm]{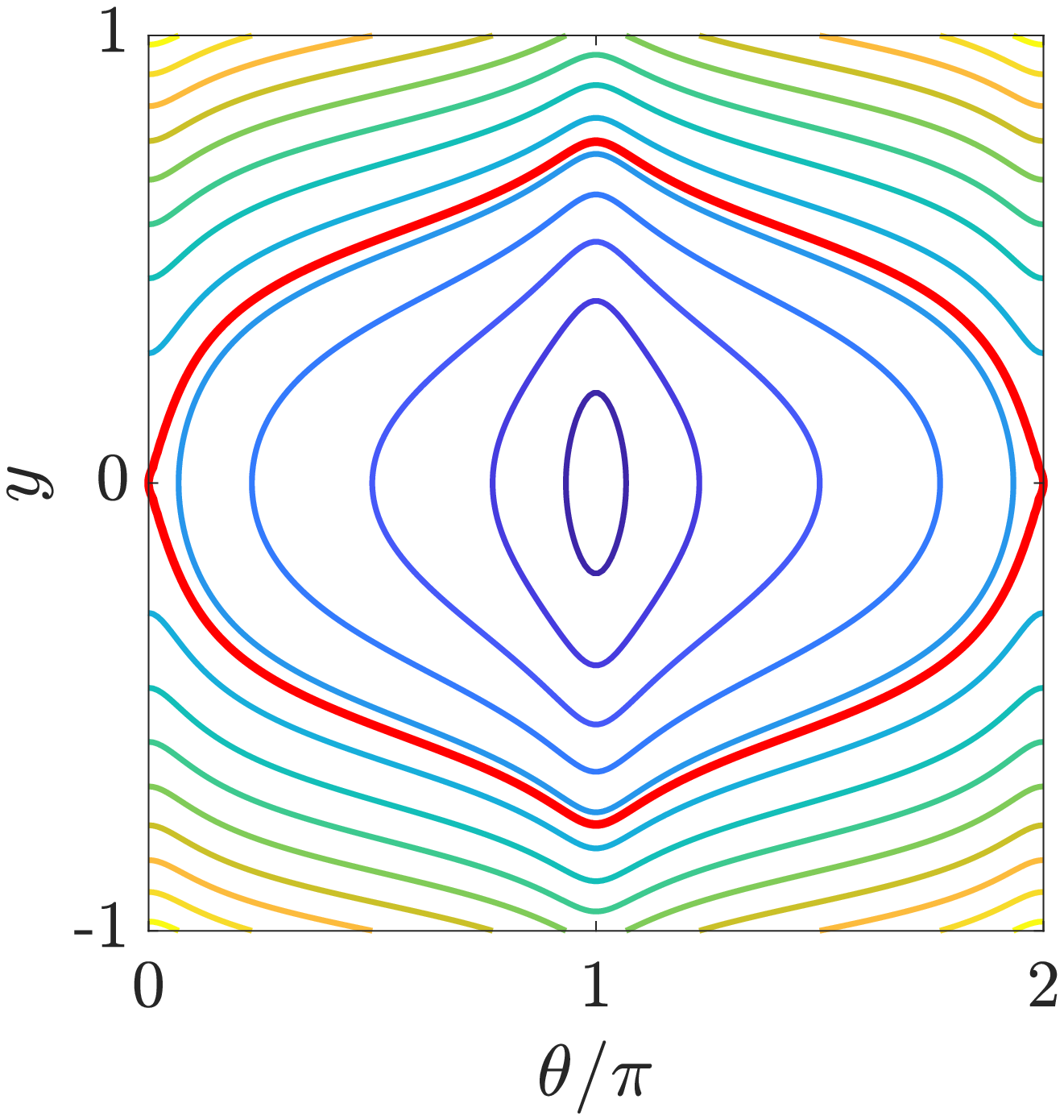}
\put(-6,95){{(b)}}
\end{overpic}
\hspace{0.2cm}
\vspace{-0.5cm}
\caption{ (Color online) Contour plots of the constants of motion in the $(\theta-y)$-phase space. (a) Spherical swimmer ($B=0$) and (b) elongated swimmer ($B=0.9$), where the flow strength is set as $\gamma=8$ and the length scale and timescale are non-dimensionalized by $h=V=1$. The separatrices are shown with thick red lines.}
\label{fig:cont}
\end{center}
\end{figure}

The system above is closed in $y$ and $\theta$, and the constant of motion is obtained by directly solving Eqs. \eqref{eq:P01b} and \eqref{eq:P01c} \cite{zottl2013periodic} as
\begin{equation}
H=\frac{\gamma}{2V}y^2+g(\theta; B)
    \label{eq:P02},
\end{equation}
where the function {\bl $g$} is given by
\begin{equation}
g(\theta; B)=\tanh^{-1}\left(\sqrt{\frac{2B}{1+B}} \cos\theta\right)/\sqrt{2B(1+B)}
    \label{eq:P03}
\end{equation}
for $B>0$.
When $B=0$, $H$ becomes $H=\frac{\gamma}{2}y^2+\cos\theta$, and this coincides with the Hamiltonian of the pendulum equation \cite{zottl2012nonlinear}. The form \eqref{eq:P03} holds even for an oblate swimmer with $B<0$ if an appropriate branch is taken. The motion in the $\theta-y$ phase plane is thus confined to the contour line of $H$, which is shown in Fig. \ref{fig:cont} for a spherical swimmer $(B=0)$ and an elongated swimmer $(B=0.9)$ with the flow strength $\gamma=8$. The behaviors of the swimmers are characterized by inner and outer regions divided by the separatrices shown with red thick curves in Fig. \ref{fig:cont}. The swimmer in the inner region faces upstream with oscillation within the chamber (``swinging"), whereas the swimmer monotonically rotates in one direction (``tumbling") in the outer region.

If the swimmer moves out of the flow plane, the dynamics are then characterized by two constants of motion $\kappa_1(\theta, \phi; B)$ and $\kappa_2(\theta, \phi; B)$, where $\phi$ is the angle made by the swimmer orientation $\bm{d}$ and the $\bm{e}_1\bm{e}_2$ plane \cite{zottl2013periodic}. The motions in the 3-dimensional $\theta-\phi-y$ space are confined to the intersections of the isoplanes of $\kappa_1$ and $\kappa_2$. Similar analyses were made for a microswimmer in a pipe with an ellipsoidal cross section, but the constant of motion was only obtained for a spherical swimmer in a cylindrical pipe  \cite{zottl2013periodic}.

\begin{figure*}[htbp]
\begin{center}
\begin{overpic}[width=17cm]{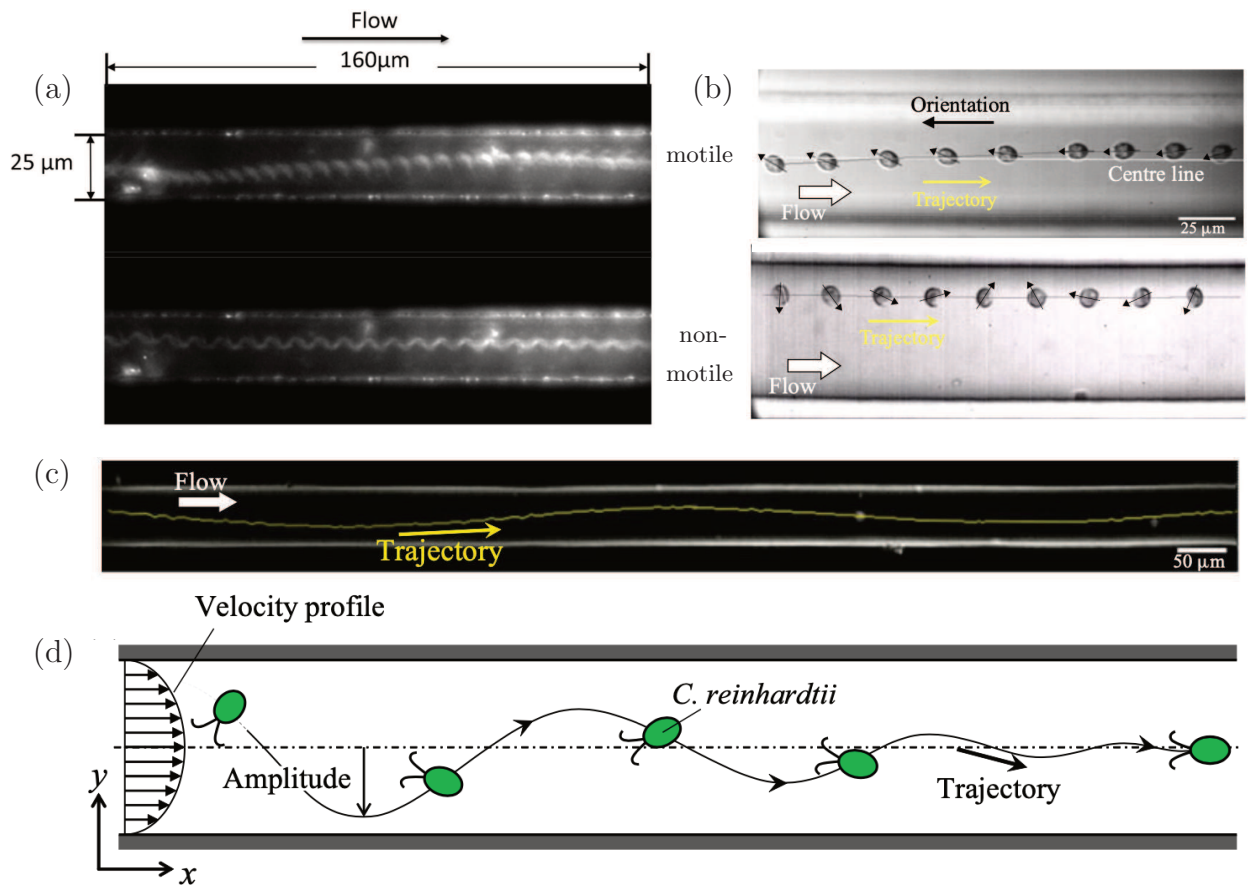}
\put(3,62){{\large (a)}}
\put(52,57){{motile}}
\put(53,43){{non-}}
\put(52,40){{motile}}
\put(54,62){{\large (b)}}
\put(3,32){{\large (c)}}
\put(3,18){{\large (d)}}
\end{overpic}
\caption{ (Color online) Experimental observations of microswimming in Poiseuille flow. (a) Superimposed snapshots of motile and immotile {\it T. brucei} cells in a microfluidic channel, in which flow is induced from left to right. A motile cell exhibits a swinging migration (upper panel) with its position moving from left to right. A non-motile cell, however, does not move in the direction perpendicular to the flow. Figure adapted from Uppaluri et al. \cite{uppaluri2012flow} with permission (copyright 2012 by Elsevier). (b--d) Sample trajectories of motile and non-motile {\it C. reinhardtii}  cells in a microchannel. The rightward-plane Poiseuille flow is generated through a microfluidic device. Figures adapted from Omori et al. \cite{omori2022rheotaxis} with permission under the creative commons license, http://creativecommons.org/licenses/by/4.0 {\rd (copyright 2022 by The Author(s)) }. (b) Superimposed snapshots of a motile {\it C. reinhardtii} cell that turns upstream (upper panel), whereas the non-motile cell only rotates with a constant height in the chamber (lower panel). (c) Sample trajectory of the movement of the cell.  (d) Schematic of trajectory and orientation of a motile {\it C. reinhardtii} cell.
}
\label{fig:pois_exp}
\end{center}
\end{figure*} 
The motility-induced swinging behavior was experimentally confirmed by a {\it Typanosoma} cell, a motile unicellular parasite \cite{uppaluri2012flow}. In Fig. \ref{fig:pois_exp}a, superimposed snapshots of both motile and non-motile {\it T. brucei} cells are adapted from Ref. \cite{uppaluri2012flow}. The upper panel shows the oscillating swinging behavior in the microchannel under Poiseuille flow from the left to the right with its orientation heading upstream. In the lower panel, in contrast, a sample trajectory of a non-motile {\it T. brucei} cell illustrates a height relative to the channel walls, showing a tumbling motion.

Experimental studies of microswimmers in Poiseuille flow were also conducted for bacteria \cite{rusconi2014bacterial, junot2019swimming}. Rusconi et al. \cite{rusconi2014bacterial} showed a flow-induced drop in the population of {\it Bacillus subtilis} in the middle region. While flow-induced trapping in the middle of the chamber was found in the numerical simulation with negligibly small rotational Brownian noise, in real experiments, the noise caused jumping across the separatrices, leading to depletion of the population. Junot et al. used smoothly swimming mutants of {\it E. coli} bacteria to study their 3-dimensional motions in a planar Poiseuille flow. They found that the theoretical model based on the Jeffery equation can predict the swimmer dynamics but only for a short time due to rotational Brownian noise.

Similar experimental observations were performed for several motile phytoplankton. Barry et al. \cite{barry2015shear} studied the cell population of four phytoplankton species, {\it Heterosigma akashiwo}, {\it Amphidium carterae}, {\it Dunaliella tertiolecta}, and {\it  Chlamydomonas reinhardtii}, in Poiseuille flow. They reported that the first two species with an elongated flagellum for swimming exhibited a depletion in the weak-shear region and the last two species with two short flagella for breaststroke swimming, in contrast, accumulated in the middle of the channel. They concluded that the effective hydrodynamic aspect ratio determined these different behaviors and also suggested a possibility of active cell response to the accumulation behavior.

More recently, Omori et al. \cite{omori2022rheotaxis} studied the motion of {\it  C. reinhardtii} cells and observed a similar central accumulation of the cell population in Poiseuille flow (Fig. \ref{fig:pois_exp}b--d). The upper panel of Fig. \ref{fig:pois_exp}b shows the reorientation of a motile cell towards upstream with accumulation in the center of the channel. This is in contrast to the behavior of the non-motile cell shown in the lower panel, which stayed in the same position relative to the channel wall while exhibiting a tumbling motion. A sample trajectory is shown with its long-time behavior (Fig. \ref{fig:pois_exp}c) and a schematic illustration (Fig. \ref{fig:pois_exp}d).

To explain the mechanism of the swimmer's central accumulation with upstream orientation, Omori et al. \cite{omori2022rheotaxis} proposed a mechanical coupling between the flow and the unsteadiness of the swimming. Considering time-periodic deformation $V=V(t)$ and $B=B(t)$, they numerically found that hydrodynamic interactions caused the microswimmers to turn upstream and accumulate at the center of the channel. This breakdown of degenerate orbits in the shape space was then theoretically formulated by using the multiple-scale analysis of Walker et al. \cite{walker2022emergent}. 

We let $\omega$ be the beat frequency of the deformation and assume that $B$ and $V$ are periodic functions with this frequency.  Introducing the new variable as $z(t)=y(t)/\omega^{1/2}$ and rescaling the equation in the form
\begin{eqnarray}
\dot{z}=\omega^{1/2}V(\omega t)\sin\theta,~\dot{\theta}=\gamma \omega^{1/2} z(1-B(\omega t)\cos2\theta)
    \label{eq:P04},
\end{eqnarray}
Walker et al. performed a multiple-scale analysis for the natural fast timescale $T:=\omega t$ and intermediate timescale $\tau:=\omega^{1/2}t$ with formal transformations of $z(t)\mapsto z(T, \tau, t)$ and $\theta(t)\mapsto \theta(T, \tau, t)$, together with
\begin{equation}
\frac{d}{dt}\mapsto \frac{\partial}{\partial t} +\omega^{1/2}\frac{\partial}{\partial \tau}+ \omega\frac{\partial}{\partial T}
    \label{eq:P05}.
\end{equation}
The time evolution of the quantity $H$ in the slow dynamics is then considered with an expansion in an inverse series of $\omega^{1/2}$ as $H=H_0+\omega^{-1/2}H_1+\omega^{-1}H_2+\cdots$. The zeroth-order quantity is then given by
\begin{equation}
    H_0=\frac{\gamma}{2\langle V\rangle}z^2_0+g(\theta_0, \langle B \rangle),
\end{equation}
where $z_0=z_0(\tau, t)$ and $\theta_0=\theta_0(\tau, t)$ are independent of the fast timescale and the bracketing $\langle ~\rangle$ denotes the average over the fast timescale $T$. The time evolution with leading-order correction is then obtained as
\begin{equation}
\frac{d H_0}{dt}=\gamma f(H_0) W
    \label{eq:P06}
\end{equation}
with a non-positive function $f(H_0)$ that depends on the swimmer's speed and shape only via their fast-time averages $\langle V\rangle$ and $\langle B\rangle$. 
The quantity $W$ is defined as
$W=\langle I (B- \langle B\rangle ) \rangle$, where
$I(T)=\int_0^T V(\tilde{T})-\langle V\rangle\,d\tilde{T}$.  Thus, if either the swimmer shape or swimmer speed is constant in time, $W$ vanishes and $H_0$ remains a constant of motion, preserving the periodic orbits in the phase space.
 When $W\neq 0$, however, $H_0$ monotonically increases or decreases depending on the sign of $W$. For swimming {\it C. reinhardtii}, due to a finite phase shift between the time oscillation of $V$ and $B$, $H_0$ is indeed decreasing in time, and therefore the swimming orientation and position are attracted by the point of $(\theta, y)=(\pi, 0)$, being oriented against the flow at the channel center. 

Unsteady swimming is not the only factor that breaks the degenerate periodic orbit. One possible factor is the effect of inertia, and Choudhary et al. \cite{choudhary2022inertial} numerically studied a spherical active particle in Poiseuille flow and showed that inertia breaks the conservational structure and leads to bifurcation in the phase space. Another factor is the hydrodynamic wall interaction, which depends on the swimming pattern of the swimmer as well as its shape.  Z\"ottl \& Stark \cite{zottl2012nonlinear} found a puller-type spherical swimmer with a constant velocity and stresslet accumulates at the channel center with its orientation facing upstream. A similar central focus for a puller swimmer was reported in numerical simulations of spheroidal squirmers \cite{qi2020rheotaxis, liu2022migration}, while these simple time-averaged parameterizations need to be derived from faster-timescale temporal deformation for a largely shape-changing
swimmer \cite{walker2022multi}. 

For a spherical swimmer, the constant of motion $H$ is indeed a Hamiltonian and the swimmer's unsteadiness may cause Hamiltonian chaos \cite{chacon2013chaotic}. A similar Hamiltonian structure is found in the 2-dimensional swimmer model in a simple shear near a planar wall \cite{ishimoto2017dynamics}, and the swimmer unsteadiness leads to Hamiltonian chaos \cite{ishimoto2017guidance}. More generally, a swimmer exhibits a large deformation, and such swimming dynamics in Poiseuille flow are studied for a sperm cell with consideration of the elastohydrodynamic coupling \cite{kumar2019effect} and for a general deforming particle \cite{tarama2017swinging}.

The conserved quantity may be broken by the fluid rheology as discussed for Jeffery's orbits. Mathijssen et al. \cite{mathijssen2016upstream} considered a spherical swimmer in a channel flow with non-Newtonian fluid rheology and found that the viscoelastic normal stress difference loses the constancy of the motion and reorients the swimmers upstream, where they accumulate at the centerline of the channel.

\subsection{More Complex Flows}

\begin{figure}[htbp]
\begin{center}
\begin{overpic}[width=8.5cm]{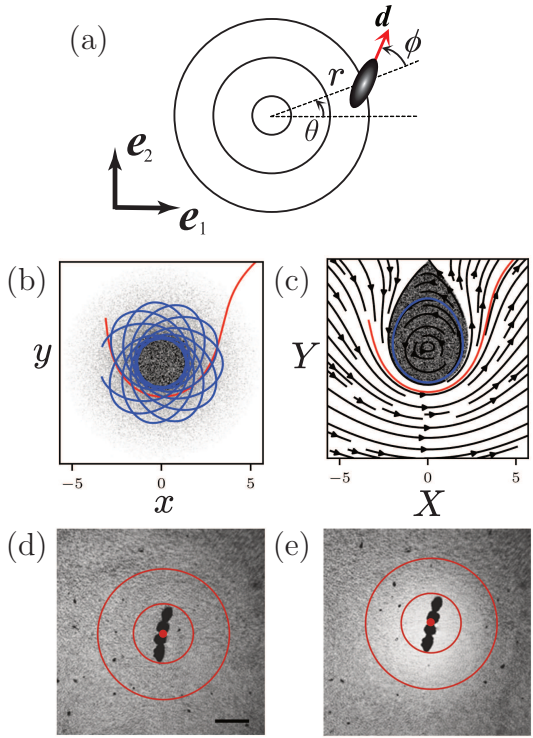}
\put(13,90){{\large (a)}}
\put(5,59){{\large (b)}}
\put(39,59){{\large (c)}}
\put(5,25){{\large (d)}}
\put(39,25){{\large (e)}}
\end{overpic}
\caption{ (Color online) Swimming in vortical flows. (a) Schematic of a microswimmer swimming in an axisymmetric vortical flow. (b,c) Theoretically obtained particle positions a long time after initially uniform distribution inside the central circle of radius 5. Selected trajectories are also shown for a trapped cell (blue) and escaping cell (red). Figure adapted from Arguedas-Leiva \& Wilczek \cite{arguedas2020microswimmers} under the creative commons license, http://creativecommons.org/licenses/by/4.0 {\rd (copyright 2020 by The Author(s)) }. (b) Plots in the laboratory frame $(x, y)$. (c) Plots in the co-rotating frame $(X, Y)$. (d,e) Depletion of {\it B. subtilis} bacteria around a rotating magnetized chain. Figure adapted from Sokolov \& Aranson \cite{sokolov2016rapid} under the creative commons license, http://creativecommons.org/licenses/by/4.0 {\rd (copyright 2016 by The Author(s)) }. (d) Initial distribution of bacteria. (e) Distribution after the depletion zone is created. White corresponds to a lower concentration.}
\label{fig:vorti}
\end{center}
\end{figure}

We then move to a more complex flow pattern. In nature, oceanic plankton typically live in a turbulent flow. The minimum size of eddies is characterized by the Kolmogorov length scale, which is a few millimeters in length in the upper mixed layer in the ocean \cite{guasto2012fluid}. Thus, most microswimmers experience unsteady shear flow. 

As a theoretical model of the smallest eddy in the environment, a swirling flow may be the simplest flow model.
Arguedas-Leiva \& Wilczek \cite{arguedas2020microswimmers} studied a microswimmer in a 2-dimensional axisymmetric vortical flow (Fig. \ref{fig:vorti}a). The background velocity field is simply given by $\bm{u}^\infty=(u(r)\cos\theta, u(r)\sin\theta)$, where $r=\sqrt{x^2+y^2}$. For the motion of a microswimmer with a constant velocity $V$ in the direction of the swimmer orientation, we employ the Jeffery equation of an axisymmetric object, assuming the flow is locally linear. 

When the swimmer shape is spherical ($B=0$), Arguedas-Leiva \& Wilczek \cite{arguedas2020microswimmers}  found a Hamiltonian structure in the swimmer dynamics with co-rotating coordinates, $X=x \cos\theta +y\sin\theta$ and $Y=-x\sin\theta+y\cos\theta$, as $\dot{X}=\partial_Y H(X, Y)$ and $\dot{Y}=-\partial_X H(X, Y)$ with a Hamiltonian $H(X, Y)=VY+\int^r sf(s)\,ds$. 
For Fig. \ref{fig:vorti}b and c, we adapted the plots in Arguedas-Leiva \& Wilczek \cite{arguedas2020microswimmers} to show the swimmer distributions after a long time in the laboratory frame (Fig. \ref{fig:vorti}b) and in the co-rotating frame (Fig. \ref{fig:vorti}c), assuming a Gaussian vortex, also known as the steady Lamb-Oseen vortex, for the background vortical flow. Initially, swimmers are uniformly distributed within a circle of radius $r=5$ whose center coincides with the origin. As shown in Fig. \ref{fig:vorti}b, some swimmers are confined near the origin in the laboratory frame, as shown by the concentration given with a grey scale. Sample swimmer trajectories are also plotted with blue and red curves, corresponding to the confined and escaping cells, respectively. This confinement is captured by the inner region separated by the homoclinic orbit in the co-rotating frame in Fig. \ref{fig:vorti}c.

While the Hamiltonian structure is broken for a non-spherical swimmer ($B\neq 0)$, the swimmer positions are still separated by the homoclinic orbits (or separatrix) in the $X-Y$ plane, demonstrating a similar confinement of microswimmers in a vortical flow. A homoclinic orbit can emerge by a saddle-node bifurcation for a small effective velocity corrected by the Bretherton parameter given by $V_{\textrm{eff}}=V/(1-B)$. When rotational noise is added, the cells escape from the trapping region around the saddle point.

Focusing on a prolate microswimmer $(B > 0)$, Tansijevi\'c \& Lauga 
\cite{tanasijevic2022microswimmers} found a constant of motion in the system of microswimming in a vortical flow generated by 3- and 2-dimentional rotlets, corresponding to the flow around a rotating sphere and cylinder, respectively. The flow field is explicitly given by $u(r)=a^3\omega/r^2$ and $u(r)=a^2\omega /r$ for the 3- and 2-dimensional rotlets, respectively, where $a$ is the radius of the sphere or cylinder and $\omega$ is the angular velocity of the rotation.

Motions of {\it B. subtilis} bacteria around a rotating object were experimentally studied by Sokolov \& Aranson \cite{sokolov2016rapid}, who reported a depletion zone around the object. In Fig. \ref{fig:vorti}d and e, we show pictures of bacteria concentrations before and after the depletion zone was formed. This depletion is well explained by the vortical confinement in numerical computation with the rotational noise of the cells \cite{tanasijevic2022microswimmers}.
This confinement of microswimmers by a vortical flow was also studied in a dumbbell model \cite{zheng2020rototaxis}, a bead-spring system \cite{kuchler2016getting}, and 
an abstract deforming particle model \cite{tarama2014deformable}.

\begin{figure}[htbp]
\begin{center}
\begin{overpic}[width=8.5cm]{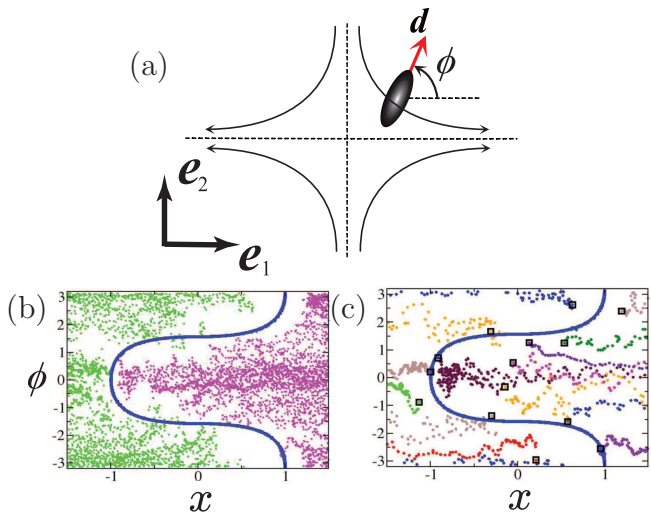}
\put(18,70){{\large (a)}}
\put(-1,32){{\large (b)}}
\put(49,32){{\large (c)}}
\end{overpic}
\caption{ (Color online) Swimming in hyperbolic flows. (a) Schematic of a swimming microswimmer in a hyperbolic flow. (b,c) Sample trajectories of smoothly swimming {\it B. subtilis} bacteria projected on the $x-\phi$ phase space, together with the swimmer invariant manifold shown with a thick blue line. Figure adapted 
with permission from Berman et al. \cite{berman2021transport}  (copyright 2021 by the American Physical Society). (b) Leftward-escaping and rightward-escaping trajectories are plotted in green and magenta, respectively. (c) Selected trajectories of each cell are shown in a different color with their initial positions being marked by open squares.}
\label{fig:hyp}
\end{center}
\end{figure}


The existence of a saddle point in the dynamical system is the key structure for the escaping behaviors of swimmers. Berman et al. \cite{berman2021transport} studied the 2-dimensional dynamics of microswimmers in a hyperbolic flow (Fig. \ref{fig:hyp}a), which is the linear flow around a saddle point with its flow field given by $\bm{u}^\infty=(Ax, -Ay)$, where the flow strength is $A>0$. The swimmer moves in the direction of the body axis, denoted by $\bm{d}$, with a constant velocity $V$. We can non-dimensionalize the system by $A=V=1$.

Passive particles are transported along the streamline. The trajectories of active particles are modified but the swimmers are still swept towards the $+x$ or $-x$ direction. The paths in the $x-y-\phi$ phase space are therefore separated by a surface, across which the sign of the final $x$ position is reversed. This surface, termed a swimmer invariant manifold, acts as a separatrix. In Fig. \ref{fig:hyp}b and c, its projection to the $x-\phi$ plane is shown with a thick blue curve. Fig. \ref{fig:hyp}b and c show the sample trajectories experimentally obtained for smoothly swimming {\it B. subtilis} cells in a hyperbolic flow taken from Berman et al. \cite{berman2021transport}. The leftward and rightward swimmers are shown in green and magenta plots in Fig. \ref{fig:hyp}b, respectively, and are well separated by the swimmer invariant manifold.  Selected trajectories in Fig. \ref{fig:hyp}b are also shown with their initial positions marked by open squares.

Hyperbolic flow induces contractile and extensile forces on the swimmer, and flexible swimmers such as sperm cells modulate their waveform during their transport \cite{kumar2019flow}. 
Similar compression is found in a flexible helix in a converging flow \cite{chakrabarti2020flexible}. Motions of rigid rod swimmers have been studied in a similar converging chamber \cite{potomkin2017focusing}.



The coexistence of confinement and escape in swimmer dynamics is more clearly shown in a 2-dimensional vortical flow lattice. In a 2-dimensional incompressible flow, passive tracers follow a stream function $\Psi(x, y)$ given by $\bm{u}^\infty=(\partial_y\Psi, -\partial_x\Psi)$. This Hamiltonian structure conserves the volume of the phase space, and thus the distribution of passive particles is homogeneous. For active particles, however, this does not hold. 

\begin{figure}[htbp]
\begin{center}
\begin{overpic}[width=8.5cm]{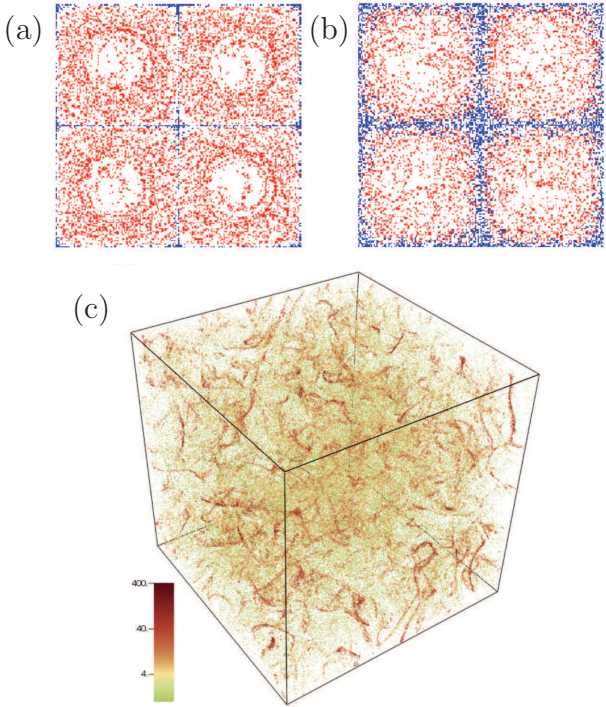}
\put(0,91){{\large (a)}}
\put(40,91){{\large (b)}}
\put(9,54){{\large (c)}}
\end{overpic}
\caption{ (Color online)  (a,b) Long-term asymptotic distribution of elongated microswimmers in a vortical lattice flow. The cells are initially uniformly distributed. With cell motility, the population separates into swimmers trapped inside the vortical cells (plotted in red) and swimmers moving across the vortical cells along their edges (plotted in blue). Figure adapted with permission from Torney \& Neufeld \cite{torney2007transport}   (copyright 2007 by the American Physical Society). (c) Snapshot of the distribution of swimmers in homogeneous turbulence. The color bar indicates the concentration of cells.
Figure adapted from Cencini et al. \cite{cencini2019gyrotactic} with permission (copyright 2019 by Springer-Nature). }
\label{fig:turb}
\end{center}
\end{figure}

Torney \& Neufeld \cite{torney2007transport} analyzed the microswimmer dynamics in vortical arrays whose stream function is given by $\Psi=(U_0L/[2\pi])\sin(2\pi x/L)\sin(2\pi y/L)$, where $L$ is the size of the lattice and $U_0$ is the maximum velocity in a vortical lattice. The orientation dynamics again satisfy the Jeffery equation, under the assumption that the background flow is locally approximated by a linear flow.
An elongated swimmer in a cellular flow can move across vortical cells if it swims with a speed higher than a critical value, and the population is then separated into two different groups: one is accumulating inside a vortical cell and the other moves around cells near the heteroclinic orbits (or edges of vortical cells). Fig. \ref{fig:turb}a and b show the long-time asymptotic distribution of particles initially distributed uniformly in space, with the two panels corresponding to different shapes and flow strength ($B=0.9$ in Fig. \ref{fig:turb}a and $B=0.7$ in Fig. \ref{fig:turb}b). The trapped swimmers and non-trapped swimmers are plotted in red and blue, respectively. 

Further analysis from the dynamical systems point of view was performed for vortex lattice and hyperbolic flows in Ref. \cite{berman2020trapping}.
The dynamics in the vortex arrays were also studied with fluid viscoelasticity \cite{ardekani2012emergence}, for gyrotactic swimmers \cite{durham2011gyrotaxis}, and for sedimenting cells \cite{clifton2018enhanced}.

Studies of microswimmer dynamics in oscillating cellular flows \cite{torney2007transport, khurana2011reduced, khurana2012interactions} suggest a patchy distribution of the cells. Indeed, formations of patches are found in gyrotactic microalgae, {\it H. akashiwo}, in vortical flows   \cite{durham2013turbulence}. Accumulation was also found for a gyrotactic model swimmer and elongated swimmer in Navier-Stokes turbulence \cite{durham2013turbulence, zhan2014accumulation, santamaria2014gyrotactic, cencini2019gyrotactic, borgnino2022alignment}. For readers interested in this topic, refer to the recent reviews \cite{cencini2019gyrotactic, qiu2022review}. Fig. \ref{fig:turb}c is a snapshot of the distribution of swimmers in homogeneous turbulence, where colors indicate the concentration of cells.


Complex flows that microswimmers face are also ubiquitous near external boundaries. Examples include a sinking sphere \cite{slomka2020encounter}, obstacles fabricated within microfluidic devices \cite{yazdi2012bacterial, dehkharghani2019bacterial}, and surface gravity waves \cite{ma2022reaching}.

\section{Summary and Outlook}
\label{sec:disc}

The Jeffery equation can be applied to many other interesting problems for exploring the effect of shape on interactions of microswimmers with external environments.

\subsection{Collective Behavior}

For modeling and simulating the collective behavior of microswimmers, the background flow for individual swimmers is approximated by summations of flow fields generated by other neighbors. In a dilute population, the far-field flows are represented by the flow induced by swimmer stresslets \cite{lauga2020fluid}. This method is also used for hydrodynamic interaction for a swimmer near a boundary, as the flow by its mirror image is taken as a background flow for the swimmer \cite{spagnolie2012hydrodynamics}. In either case, the swimmer's shape affects whether the interactions are attractive or repulsive.

These interactions have been scaled up to collective dynamics \cite{saintillan2008instabilities, reinken2018derivation} and used in modeling and understanding bioconvection patterns \cite{bees2020advances} and active turbulence \cite{alert2022active}. The orientation dynamics are also significant in the rheological properties of suspensions of microswimmers\cite{saintillan2013active, saintillan2018rheology}. In a dense suspension, however, the swimmer shape is also significant due to steric and near-field hydrodynamic interactions \cite{kyoya2015shape, bar2020self}. 

\subsection{Smart Swimmers}

Control and navigation of a microswimmer under a flow are also active topics. What is the optimal strategy for a swimmer that can sense the flow around it \cite{monthiller2022surfing}? How can we optimize the flow field to control a swimmer's path inside a microdevice \cite{moreau2021control, moreau2021driving}? Recently, adaptive and tactic swimming behaviors in {\it smart swimmers} have attracted remarkable attention in the context of creating and designing such artificial smart swimmers \cite{tsang2020roads} and understanding adaptive behaviors of real biological swimmers in complex environments
\cite{sengupta2017phytoplankton, ishikawa2022bacterial, echigoya2022switching}, where the 
use of machine learning is rapidly emerging, for example, in active matter physics \cite{cichos2020machine} and fluid mechanics \cite{brunton2020machine}.

In particular, reinforcement learning is used to obtain an optimal swimming stroke \cite{tsang2020self, zou2022gait}, optimal flow mixing \cite{konishi2022fluid}, and an optimal  locomotion strategy \cite{muinos2021reinforcement}. Many studies have recently been performed on navigation of a microswimmer through a complex flow environment, such as Taylor-Green vortical cells \cite{colabrese2017flow}, a complex potential \cite{schneider2019optimal}, a chaotic ABC flow \cite{gustavsson2017finding}, in Navier-Stokes turbulence \cite{alageshan2020machine}, and in a fluid with spatially varying viscosity \cite{nasiri2022reinforcement}, in addition to the gyrotactic stability problem \cite{qiu2022active} and predation dynamics \cite{borra2022reinforcement, zhu2022optimizing}.

\subsection{Concluding Summary}

In this review, we have provided a theoretical introduction to microswimmer hydrodynamics, the dynamics of a self-propelled particle in a low-Reynolds-number flow, with an extensive focus on the Jeffery equations that govern the orientation dynamics of a particle in a linear flow. By introducing the concept of {\bl hydrodynamic} symmetry, we have discussed the theoretical backgrounds of the Jeffery equations and their extension to objects with more general symmetry and objects that may rapidly deform (Sec. \ref{sec:ext}). The importance and usefulness of the Jeffery equation are also discussed through its periodic motion in simple shear, which emerges as a constant of motion. Motions of microswimmers in simple shear flow, Poiseuille flow, vortical flow, hyperbolic flow, and other complex flow patterns are summarized in Sec. \ref{sec:flow}, with special attention to the emergence and breakdown of the constants of motion in different situations. Since its first derivation by Jeffery a century ago, the motions of non-spherical particles and swimmers in low-Reynolds-number flows have opened up extensive research fields in material, life, and engineering sciences, and at the same time have provided a large variety of non-linear dynamics solutions from simple equations.

\vspace{3em}





K.I. acknowledges support from the Japan Society for the Promotion of Science (JSPS), KAKENHI for Young Researchers (Grant No. 18K13456); JSPS, KAKENHI for Transformative Research Areas (Grant No. 21H05309); and the Japan Science and Technology Agency (JST), PRESTO (Grant No. JPMJPR1921). {\bl K.I. acknowledges M. P. Dalwadi, E. A. Gaffney, C. Moreau, and B. J. Walker for stimulus discussions.} 
This study was partially supported by the Research Institute for Mathematical Sciences, an International Joint Usage/Research Center located at Kyoto University.




\bibliography{library.bib}

\end{document}